\documentstyle[aps,prb,multicol,epsf,graphicx,amsmath]{revtex}

\newcommand \bc {\begin{center}}
\newcommand \ec {\end{center}}
\newcommand \ee {\end{equation}}
\newcommand \be {\begin{equation}}
\newcommand \beq {\begin{eqnarray}}
\newcommand \eeq {\end{eqnarray}}
\newcommand \bmu {\begin{multline}}
\newcommand \emu {\end{multline}}
\newcommand \e {\epsilon}
\newcommand \eps {\epsilon}
\newcommand \nn {\nonumber}
\newcommand{\eq}[1]{(\ref{#1})}
\newcommand{\eqq}[2]{(\ref{#1},\ref{#2})}
\newcommand{\eqqq}[3]{(\ref{#1},\ref{#2},\ref{#3})}

\begin{document}

\title{Mode-Dependent Nonequilibrium Temperature in Aging Systems}
\author{A.Garriga$^1$ and  F. Ritort$^2$\\
$^1$ Departament de Tecnologia\\Universitat Pompeu Fabra\\
Passeig de Circumval.laci\'o 8, 08003 Barcelona (Spain)\\
$^2$ Departament de Fisica Fonamental, Facultat de F\'{\i}sica\\
Universitat de Barcelona\\
Diagonal 647, 08028 Barcelona (Spain)\\
E-Mail: adan.garriga@upf.edu, ritort@ffn.ub.es}
\date{\today}
\maketitle

\begin{abstract}
We introduce an exactly solvable model for glassy dynamics with many
relaxational modes, each one characterized by a different relaxational
time-scale. Analytical solution of the aging dynamics at low
temperatures shows that a nonequilibrium or effective temperature can be
associated to each time-scale or mode. The spectrum of effective
temperatures shows two regions that are separated by an age dependent
boundary threshold. Region I is characterized by partially equilibrated
modes that relax faster than the modes at the threshold
boundary. Thermal fluctuations and time-correlations for modes in region
I show that those modes are in mutual thermal equilibrium at a unique
age-dependent effective temperature $\Theta (s)$. In contrast, modes
with relaxational timescales longer than that of modes at the threshold
(region II) show diffusive properties and do not share the common
temperature $\Theta (s)$. The shift of the threshold toward lower energy
modes as the system ages, and the progressive shrinking of region II,
determines how the full spectrum of modes equilibrates. As is usually
done in experiments, we have defined a frequency-dependent effective
temperature and we have found that all modes in region I are mutually
equilibrated at the temperature $\Theta (s)$ independently of the
probing frequency. The present model aims to explain transport anomalies
observed in supercooled liquids in terms of a collection of structurally
disordered and cooperative rearranging mesoscopic regions.

\end{abstract}

\section{INTRODUCTION}
\label{intro}

Finding a general theory for nonequilibrium systems constitutes one of
the major challenges in modern physics. Despite of the many different
existing approaches to nonequilibrium phenomena
\cite{des_mazur,des_kubo}, at present there is not yet a general
theory applicable to systems arbitrarily far from equilibrium. During
many decades, physicists have studied different sorts of
nonequilibrium systems. A common strategy has been to adapt well-known
thermodynamic concepts to the nonequilibrium regime with the hope that
the most fundamental assumptions still hold in a varied range of
nonequilibrium conditions. One of the concepts that physicists have
tried to rescue repeatedly is that of local equilibrium
\cite{des_rubivilar} and the possibility of defining a nonequilibrium
temperature \cite{des_jou}. Different definitions of a nonequilibrium
temperature have been explored for nonequilibrium liquids
\cite{EvansMorriss}, steady-state driven systems
\cite{sasatasaki,des_kurreo}, granular media \cite{des_edwards} and
glassy systems \cite{des_kurchan,des_grigera}. Admittedly, the concept
of a nonequilibrium temperature in realistic systems still remains
elusive, mainly because of the scarce number and difficulty of the
experiments that have to address fundamental predictions by the theory
\cite{des_cil2,des_felixnoeq1}.

Compelling results have been obtained over the past two decades in the
context of glassy systems. Several concepts (e.g. mutual
equilibration) have been rationalized by borrowing mathematical tools
used in the context of spin glasses (such as reparametrization-time
invariance or supersymmetry \cite{des_kurnature}). Glassy systems are
non-stationary systems which relax very slowly toward equilibrium
\cite{des_sit}. Relaxational processes can be characterized by the
\textit{age} of the glass, which is the time elapsed since the system
was prepared in the non-equilibrium state. A key signature of aging is
the fact that correlations and responses decay in a time-scale which
is roughly proportional to the age of the system, thereby showing
strong and complicated hysteretic effects\cite{des_bouch}. Theoretical
studies in mean-field spin-glasses have shown\cite{des_kurchan} that a
nonequilibrium temperature can be defined in the aging regime. This
entails an extension of the fluctuation-dissipation theorem (FDT) to
the aging regime. This nonequilibrium temperature (usually called {\em
effective temperature} in the glass jargon) is defined by,
\be k_B T_{\textrm{eff}}(t,s) =\left(\frac{\frac{\partial
C(t,s)}{\partial s}}{G(t,s)} \right)~~~~~ t>s
~~~~~~,\label{des_teff}\ee
where $C(t,s)$ is a generic two-times correlation function and
$G(t,s)$ is the corresponding response of the system to an external
perturbation applied at a given previous time $s$. At first glance,
\eq{des_teff} is just a useful way to quantify deviations from thermal
equilibrium (the equilibrium regime is characterized by
$T_{\textrm{eff}}(t,s)=T$).

In experiments \cite{des_cil2} the effective temperature is usually
defined in terms of quantities measured in frequency space
\cite{des_felixsoll}:
\be k_B T_{\textrm{eff}}^F(\Omega,s)=\Omega
\frac{S(\Omega,s)}{\widehat{\chi}''(\Omega,s)}~~~~,
\label{des_teffF}
\ee
where $\Omega$ is the probing frequency, $S(\Omega,s)$ is the power
spectrum of fluctuations at time $s$ and $\widehat{\chi}''(\Omega,s)$
is the dissipative part of the response. In equilibrium
$T_{\textrm{eff}}=T$ independently of the probed frequency $\Omega$.

Over the past years the validity of this parameter from a statistical
and thermometric point of view has been widely studied
\cite{des_cugli,des_felixcris}. Yet, there are still several debated
issues. Of the utmost importance is to understand under which
conditions the effective temperature governs the heat flow between
different aging systems, whether a zero-th law is applicable between
different systems sharing a common effective temperature
\cite{des_art4} and whether such temperature is experimentally
measurable \cite{des_felixnoeq1}. More speculative but not less
important is whether we could use such measurements to trace back the
age of a man-made glass, i.e. to approximately determine the time
elapsed since it was quenched. Our capability to answer such a
question might help us to understand the limitations and difficulties
in the experimental measurement of effective temperatures.

No doubt that many aspects related to the nonequilibrium properties of
glasses are tightly dependent on the microscopic details of the system
under study (a polymer blend, a structural glass former, a disordered
magnet or a dirty superconductor). However, generic and universal
aspects of the nonequilibrium thermal properties in glassy systems are
expected beyond the specific properties of the system under
study. This has motivated the appearance over the past decades of
several phenomenological approaches to the glassy state, from the old
free volume theories to the Adam-Gibbs-DiMarzio entropic
theories~\cite{AdamGibbs65}, the free-energy landscape
approaches~\cite{Goldstein69} until the more recent mosaic
pictures~\cite{XiaWol,BouBir}.

A major question that has recently attracted much attention in
experimental studies of supercooled liquids is the microscopic origin
of the non-exponential behavior observed in the relaxation
functions. According to some views, a glass is spatially homogeneous
and the relaxation of local regions in the system is intrinsically
non-exponential.  Another view supports the idea of a glass as an
intrinsically heterogeneous system made out of a collection of
cooperative mesoscopic regions in the system, each one showing
exponential relaxation with its own characteristic relaxation
time. Ergodicity also implies that regions must have a spectrum of
finite lifetimes and that these regions intermittently switch from
mobile to frozen. The superposition of the relaxation of the many
local regions gives rise to the non-exponential behavior observed in
the relaxation of bulk quantities. Concomitant to the experiments
there has been an upsurge in the theoretical study of heterogeneities
in simple models using analytical methods
\cite{des_ricci,des_hetero,des_chand} and numerical simulations
\cite{Don98}. In general these models support the heterogeneous
scenario in agreement with what has been observed in experiments
\cite{des_ediger,des_isra,des_bout}. If the heterogeneous scenario is
valid then one can ask whether it is possible to define a mesoscopic
effective temperature locally defined for each region and
intermittently fluctuating at time intervals roughly equal to the
lifetime of the region.  In such case it seems more reasonable to talk
about a fluctuating effective temperature field rather than a single
well defined value. Recent AFM \cite{des_isra}, fluorescence
spectroscopy \cite{des_bout} and Nyquist noise measurements
\cite{des_cil2,des_cil1} seem to endorse the heterogeneous
scenario. Moreover, Nyquist noise experiments purport evidence of
frequency dependent effective temperatures \eq{des_teffF} which can be
interpreted in terms of degrees of freedom that have different
effective temperatures and not being in mutual thermal equilibrium.

Inspired by these facts, we introduce an exactly solvable model with
glassy dynamics, the Disordered Oscillator Model (DOM). The model is
an extension of a previous one introduced in Ref.\cite{des_bpr} as a
simple model for a glass with slow dynamics in the presence of entropy
(rather than energy) barriers, the so-called Oscillator Model
(OM). The OM represents a simple example of cooperative slow dynamics
due to entropy barriers where kinetics is constrained by the dynamical
rules rather than by the Hamiltonian. In the present paper we show how
a mode dependent effective-temperature can be defined in the presence
of many different time-scales. This is naturally accomplished by
adding structural disorder in the OM (a similar strategy has been
followed for the Backgammon model \cite{des_leuzzi}). In the presence
of disorder the system preserves aging dynamics at low temperatures
and a mode-dependent effective temperature which is higher than the
bath temperature drives the relaxation of the system toward
equilibrium. The scenario of relaxation in this model is as follows:
relaxational modes can be split in two regions (I and II) separated by
a threshold boundary. Region I contains mutually thermalized modes at
an age-dependent (denoted by $s$) but frequency independent effective
temperature $\Theta (s)$ that lies above the bath temperature $T$. In
contrast, modes in region II are non-thermalized and
diffusive. Relaxation takes place by the progressive mutual
thermalization of degrees of freedom in region I to the single
temperature $\Theta (s)$. As the system ages, the threshold region
shifts, $\Theta (s)$ decreases further and more modes enter in region
I. Full thermalization occurs when $\Theta$ has relaxed to $T$ and
region II has vanished.

The paper is organized as follows. In Section~\ref{sec2} we introduce
the model. In Section~\ref{sec3} we present the formal solution of the
model. In Section~\ref{sec4} we derive the dynamical evolution for the
energy, correlations and responses. In Section~\ref{sec4} we also
analyze the behavior of the non-equilibrium effective temperatures for
the different modes in the system. In Section~\ref{sec5} we derive the
effective temperature using fluctuation relations. Section~\ref{sec6}
summarizes the results and Section~\ref{sec7} presents a discussion of
the implications of this work to our understanding of the glassy
state. Technical aspects are described in the Appendices.

\section{THE DISORDERED HARMONIC OSCILLATOR MODEL (DOM)}
\label{sec2}

\begin{figure}[hbp!]
\begin{center}
\includegraphics*[width=10cm,height=8cm]{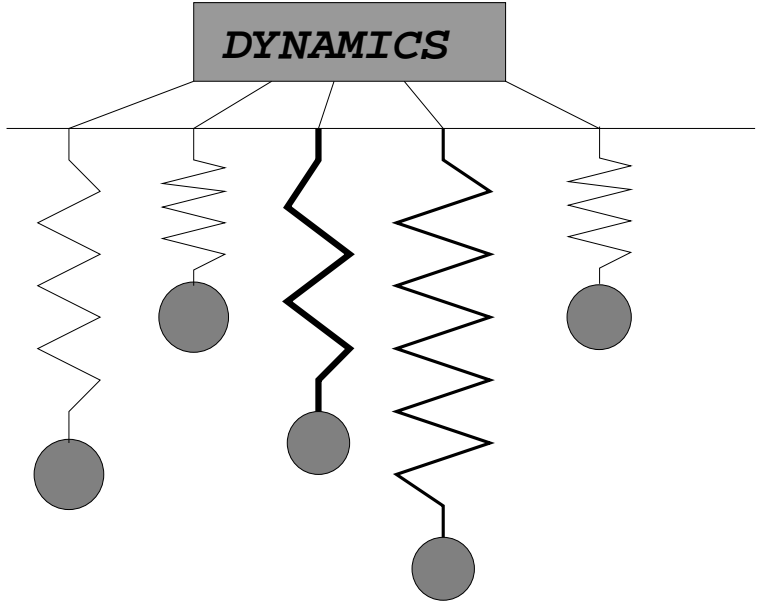}
\vskip 0.1in \caption{Schematic picture of a system of independent
oscillators with different springs constants. The thickness of the
springs represents different values for the spring constants. The
motion of the oscillators is not independent due to the coupling
induced by the Monte Carlo dynamics (represented by the top
rectangular box with the heading ``dynamics'').\label{fig1}}
\end{center}
\end{figure}

The original OM model was introduced in \cite{des_bpr} as a simple
entropic model for a glass. It consists of $N$ identical and uncoupled
harmonic oscillators with Hamiltonian:
\begin{equation}
H = \frac{1}{2} K \sum_{i=1}^N  x_i^2~~~~~\label{des_hamiltonian}
\end{equation}
where $K$ is the stiffness or spring constant and the $x_i$ are the
positions of the oscillators. The dynamics of the model is stochastic
of Monte Carlo type. A Monte Carlo trial move is defined as
follows. The position of all oscillators $ x_i $ are simultaneously
shifted by $ x_i + r_i/\sqrt{N} $ where $ r_i $ are random variables
extracted from a Gaussian distribution with zero mean and variance
$\Delta^2$. The global move is accepted according to a transition
probability $W(\Delta E) $ which satisfies detailed balance: $
W(\Delta E) = W(-\Delta E)\exp(-\beta \Delta E ), $ where $ \Delta E $
is the change in the energy \eq{des_hamiltonian} and $\beta=1/(k_BT)$
(from now on we set the Boltzmann constant equal to 1, $k_B=1$). We
use the Metropolis algorithm for the dynamics.

\be W(\Delta E) = \min [1,\exp{(-\beta \Delta E
)}]~~~~~~.\label{des_metro}\ee

Note that whenever a move is accepted, all oscillators are updated
in parallel which means that the motion of all oscillators is
cooperative and coupled by the
dynamical rule. Therefore, although they are non-interacting in the
Hamiltonian \eq{des_hamiltonian}, the dynamical rule couples all the
oscillators in a non-trivial way.

At low temperatures the OM shows typical non-equilibrium features of
glassy systems such as aging in the correlation and response
functions. Indeed, at very low temperatures the acceptance of the
Monte Carlo updates decreases very fast (like $1/t$). In this way
relaxation generates entropic barriers which make dynamics extremely
slow (the energy decreases as $1/\log t$). The simplicity of this
model, however, makes it exactly solvable. For a compilation of other
results see Refs.(\cite{des_felixsoll,des_art1,des_teo}). The OM has
also been extended to include non-harmonic generalized potentials
\cite{des_felixnoeq2}.

In the present work we add different time-scales to the system by
introducing disorder in the different degrees of freedom that enter
into the Hamiltonian. The disordered oscillator model (DOM) is defined
by the following Hamiltonian:
\begin{equation}
H_0 = \frac{1}{2} K \sum_{i=1}^N \epsilon_i
x_i^2~~~~~.\label{des_hamilt}
\end{equation}
where $K\epsilon_i$ is the spring constant of each oscillator.  As in
the original model, the system consists of a set of non-interacting
harmonic oscillators evolving in parallel according to a Monte Carlo
dynamics \eq{des_metro}. Again, the dynamics is cooperative and
couples the evolution of all oscillators. In this disordered version,
the spring constants of the oscillators have a probability density
given by,
\be g(\e) = \frac{1}{N}\sum_i \delta (\e_i-\e)~~~.\label{distrib}\ee
We assume that $\e_i> 0,\forall i$. The OM \eq{des_hamiltonian} is a
particular case of the DOM with $g(\e)=\delta(\e-1)$. In
Fig.~\ref{fig1} we show a schematic picture of the model where a set
of different oscillators move coupled according to the dynamical
rule. The inclusion of disorder in the DOM allows us to explicitly
consider nonequivalent degrees of freedom in the system having
different relaxational properties.

Thermodynamics of the DOM is trivial. The free energy per oscillator
$f$ is given by,
\be f=-\frac{1}{2}\log\bigl(\frac{2\pi}{\beta K}
\bigr)+\frac{1}{2}\int_0^{\infty}d\e g(\e)\log(\e)~~~~,
\label{free}
\ee
where $g(\e)$ is defined in \eq{distrib}. This expression is valid for
any finite $N$ if $\e_i> 0,\forall i$. In the large $N$ limit, $g(\e)$
may become a continuous function of $\e$. The last integral term in
\eq{free} can diverge depending on the behavior of $g(\e)$ when $\e\to
0$. Whatever value this last term takes, it only represents an
additive constant to the total entropy. It may be taken as defining
the entropy of a reference state without influencing the rest of
thermodynamic quantities. As we will see later, the behavior of the
oscillators with coupling constants $\e\to 0$ determines the global
relaxational properties of the system.

\section{FORMAL SOLUTION OF THE MODEL}
\label{sec3}

In this section we derive the dynamical equations for the energy and
the acceptance rate of the system. To this end we must compute the
probability distribution of having a change $\delta E$ in the energy
$E=H_0/N$ \eq{des_hamilt} in one Monte Carlo step. In a Monte Carlo
step the $N$ oscillators are moved according to the rule:
\be x_i \longrightarrow
x_i+\frac{r_i}{\sqrt{N}}~~~~\forall i~~~~,\label{des_element}
\ee
where $r_i$ are random
variables Gaussian distributed with zero mean and variance
$\Delta^2$. The change in energy is given by, \be \delta E =
K\sum_{i=1}^{N}\e_i\left(\frac{r_ix_i}{\sqrt N}
+\frac{r_i^2}{2N}\right)~~~~.\ee In what follows we identify
$\sum_{i=1}^N$ by $\sum_i$ without risk of confusion. The probability
of having a change $\delta E$ can be written as,
\beq P(\delta E) = \int_{-\infty}^\infty \delta \left( \delta E -
K\sum_i\e_i\left(\frac{r_ix_i}{\sqrt N}
+\frac{r_i^2}{2N}\right)\right) \left( \prod_i
\frac{dr_i}{\sqrt{2\pi\Delta_1^2}} \exp
\left(-\frac{r_i^2}{2\Delta_1^2}\right)\right)~~.\nn\\ \eeq
Using the integral representation of the Dirac's delta function
and performing the Gaussian integrals we obtain:
\begin{equation}
P(\delta E) = \frac{1}{\sqrt{4\pi K Q_2\Delta^2}} \exp \left( -
\frac{(\delta E - \frac{K\Delta^2 \widehat{\e}}{2})^2}{4K
Q_2\Delta^2} \right)~~~~~,\label{des_pdeltae}
\end{equation}
where we have defined the following quantities:
\be Q_k=<\frac{K}{2N}\sum_i \e_i^k x_i^2> ~,~~~
{\widehat{\e}}=\frac{1}{N}\sum_i \e_i~,\label{des_q_2} \ee where
the brackets $<..>$ denote an average over initial conditions and
dynamical histories of the system.

Note that \eq{des_pdeltae} is analogous to that obtained in the OM
\cite{des_bpr} (where $\e_i=1~,\forall i$, and $Q_k=KE, \forall k$). In
the DOM the energy is given by $E=Q_1$.  Comparing to the OM
\eq{des_pdeltae} now depends on $Q_2$ rather than on the energy $E$. As
in the case without disorder, the evolution for the energy is given by
the expression:

\be \frac{\partial E}{\partial t} = \int_{-\infty}^0 dx x P(x) +
\int^{\infty}_0 dx x P(x)\exp(-\beta x)~~~~, \ee where we have
used \eq{des_metro}. Straightforward calculations give,

\begin{equation}
\frac{\partial E}{\partial t} = \frac{a_c}{2} \left(
\textrm{erfc}(\alpha)+\left(1-\frac{
2K\Delta^2Q_2\beta}{a_c}\right) f_{Q_2}(t)
\right)~~~~~,\label{des_energy}
\end{equation}
where we have defined the following quantities
\begin{equation}
a_c=\frac{1}{2} K\Delta^2{\widehat{\e}}~,~~~~~\alpha (t)
=\frac{a_c}{\sqrt{4K\Delta^2Q_2(t)}},~~~~
\textrm{erfc}(x)=\frac{2}{\sqrt{\pi}} \int_x^\infty \exp(-y^2) dy,
\label{eqIIIerfc}
\end{equation}
\begin{equation}
f_{Q_2}(t) = \exp \left(-\beta a_c \left(1-\frac{\beta
a_c}{4\alpha^2(t)}\right) \right) \textrm{erfc}\left(\alpha (t)
\left(\frac{\beta a_c}{2\alpha^2(t)} - 1\right)\right)~.
\label{ft}
\end{equation}

It is easy to see that for $\e_i=1,\forall i$, the results of
Ref.\cite{des_bpr} are recovered. The acceptance or average fraction
of accepted moves is

\be
A(t)=\int_{-\infty}^0 dx  P(x) +
\int^{\infty}_0 dx P(x)\exp(-\beta x)~~~. \label{aceptacio}
\ee

Inserting \eq{des_pdeltae} in \eq{aceptacio} we get:

\be A(t)=\frac{1}{2}\Big( \textrm{erfc}(\alpha(t)) + f_{Q_2}(t)
\Big)~~~. \label{des_accept}\ee

To solve \eq{des_energy} and \eq{des_accept} we note
that the equations for the energy and the acceptance are not
closed because they depend on the quantity $Q_2(t)$. The equation
for $Q_2$ turns out to depend on $Q_3$ and $Q_2$. In general, the
equation for $Q_k$ depends on $Q_{k+1}$ and $Q_2$. In Appendix A we
present the dynamical equations for $Q_k$. The final
result \eq{des_jer} is:

\be \frac{\partial Q_k}{\partial t} = \left( \frac{K\Delta^2
\widehat{\e_k}}{2}-\frac{a_c Q_{k+1}}{Q_2}\right) A(t) +
\left(\frac{Q_{k+1}}{Q_2}\right) \frac{\partial E}{\partial t}~~
~~~~~~ k\geq 2 ~~~\label{des_jer2},\ee 
where we have defined the moments of the coupling distribution: 
\be {\widehat{\e_k}}= \frac{1}{N}\sum_i \e^k=\int_0^{\infty}\e^kg(\e)d\e~~.\label{moments}\ee

In what follows we will identify $\widehat{\e}=\widehat{\e_1}$. We must
notice that the hierarchy \eq{des_jer2} can be solved in the same way as
in the case of the spherical Sherrington-Kirkpatrick model
\cite{des_esferico} or the Backgammon model \cite{des_backgammon} by
introducing a generating function and solving the resulting partial
differential equation by the method of the characteristics. We will not
pursue this strategy here, rather we will analyze the simpler adiabatic
approximation.

\subsection{Adiabatic approximation: the zero-temperature limit}

Equation \eq{des_energy} can be solved at zero temperature by
means of an adiabatic approximation. It states that some quantities
evolve much slower than the energy so we can consider them as partially
equilibrated over the constant energy surface. At equilibrium the derivative of
the energy \eq{des_energy} is zero and $Q_2^{\textrm{eq}}$ is given by:
\be Q_2^{\textrm{eq}}= \frac{T}{2} \widehat{\e}=
E^{\textrm{eq}}\widehat{\e}~~~~.\label{des_q2eq}\ee
The adiabatic approximation becomes asymptotically valid for long enough
times. It relates $Q_2(t)$ with the energy $E(t)$ by
assuming that the system has partially equilibrated over the hypersurface of energy $E(t)$:
\be Q_2 (t)= E(t) \widehat{\e}~~~~\label{des_adiabatic}\ee
Within the adiabatic approximation \eq{des_energy} is
closed and can be solved analytically. In fact, this equation is
identical to the equation of the energy in the OM model. As
was done in Ref.\cite{des_bpr} the time-evolution equation for $\alpha
(t)$ \eq{eqIIIerfc} at zero temperature is:

\be \frac{\partial \alpha}{\partial t}=\frac{\exp
(-\alpha^2)}{\sqrt{\pi}}~~~~,\ee where we have expanded the error
function in the limit $\alpha\to\infty$. For large times, the
parameter $\alpha$ grows logarithmically as a function of time:
\be \alpha (t) \simeq \left( \log(\frac{2t}{\sqrt{\pi}})+\frac{1}{2}
\log(\log(\frac{2t}{\sqrt{\pi}}))\right)^\frac{1}{2}~~~~,\ee
therefore, the acceptance is independent of the disorder distribution
$g(\e)$ and decays as, 
\be A(t) \simeq
\frac{1}{4t\log(\frac{2t}{\sqrt{\pi}})}~~~~,\label{des_accept_adi}\ee
plus subdominant logarithmic corrections. The energy decays
logarithmically in time and depends only on the mean value of the
distribution, $\widehat{\e}$:

\be E(t)\simeq \frac{K\Delta^2
\widehat{\e}}{16}\frac{1}{\log(\frac{2t}{\sqrt{\pi}})+
\frac{1}{2}\log(\log(\frac{2t}{\sqrt{\pi}}))}~~~~.\label{des_energy_adi}\ee

\begin{figure}[hbp!]
\begin{center}
\includegraphics*[width=10cm,height=8cm]{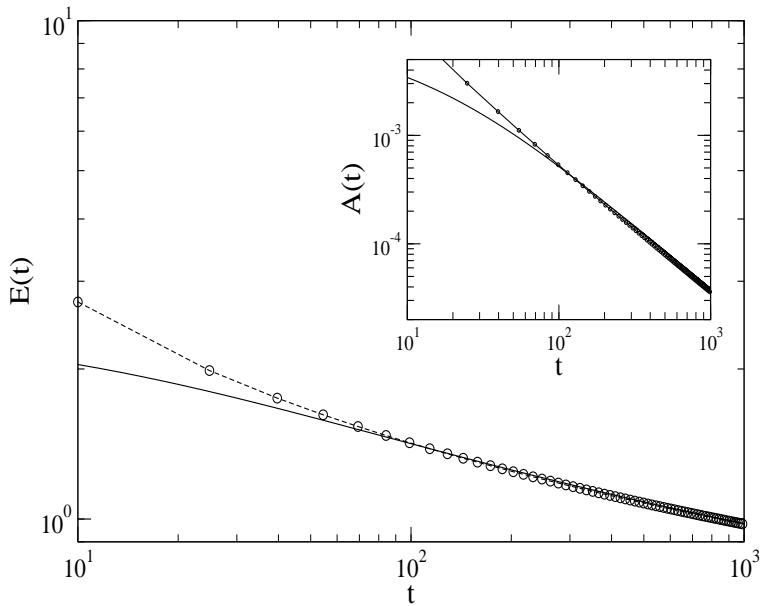}
\vskip 0.1in \caption{Zero-temperature time evolution of the energy
for a Gaussian distribution with mean $\widehat{\e}=\Delta=5$. The
time is measured in Monte Carlo steps and the energy is
dimensionless. The dashed line with circles is the asymptotic solution
\eq{des_energy_adi} while the continuous line is the numerical
solution of \eq{des_energy}. In the inset we show the acceptance. The
dashed line with circles correspond to the adiabatic solution
\eq{des_accept_adi} and the continuous line is the exact solution of
the numerical equations. To solve \eqq{des_energy}{des_accept} we have
integrated up to 500 terms in the hierarchy \eq{des_jer2}.
\label{fig2}} 
\end{center}
\end{figure}

In Fig.~\ref{fig2} we show the time evolution (measured in Monte Carlo
steps) of the energy for a Gaussian distribution of the disorder. The
dashed line filled with circles is the asymptotic result
\eq{des_energy_adi} which is in agreement with the numerical solution
of \eq{des_energy} (continuous line) for asymptotically large
times. In the inset of Fig.~\ref{fig2} we show the acceptance. Again,
circles are the asymptotic solution \eq{des_accept_adi} while the
continuous line is the numerical solution of the exact equation
\eq{des_accept}.

From \eq{des_q2eq} and the adiabatic approximation \eq{des_adiabatic} we
can define a time-dependent effective temperature (that we denote as $
\Theta(t)$) given by:
\be \Theta(t)=
\frac{2Q_2(t)}{\widehat{\e}}~~~~.\label{des_teff_adi}\ee
At this stage of the calculations this is just a definition which
denotes the temperature that we could assign to the system if it were
partially in equilibrium over the hypersurface of constant energy $E(t)$
(meaning that it reaches the maximum entropy state compatible with that
energy). In the following sections we will show how this effective
temperature drives the relaxation of some degrees of freedom in the
system. In addition, we will show that $\Theta(t)$ can be exactly derived
by assuming that energy fluctuations are described by a non-equilibrium
measure reminiscent of the Edwards measure introduced in the context of
granular media~\cite{des_edwards2}.

\section{ANALYTICAL SOLUTION OF MODE-DEPENDENT QUANTITIES}
\label{sec4}

In this section we analyze the relaxation of the different degrees of
freedom in the system. We define the mode-energy density which is the
energy density of those oscillators with a given spring constant value $K\e$,
and compute its time evolution. Then we compute the
correlations and responses for the different modes. Finally, we study
the effective temperature for the different degrees of freedom in the
system.

\subsection{MODE-ENERGY DENSITY RELAXATION}
\label{subsec1}

A physical quantity which provides deeper insight into the behavior of
the system is the \textit{mode-energy density} $w(\epsilon,t)$ defined
as follows:
\be w(\e,t)=<\frac{K}{2N}\sum_i \delta (\e_i - \e) x_i^2(t)>~~~~.
\label{des_energydens}\ee
This quantity corresponds to the mode-energy density of those
oscillators with the same spring constant $K\eps$.  The total
energy of the system can be written in terms of the mode-energy density
as:
\be E(t)=\int_0^{\infty} \e w(\e,t) d\e~~~. \ee
The characteristic relaxation time $\tau(\e)$ of a given mode is
inversely proportional to $K\e$ (see the computations in Appendix
B). Therefore, $w(\e,t)$ is the energy density of those oscillators
with a given characteristic relaxation time
$\tau(\e)$ \eq{des_eqtime}. In what follows we consider the time
evolution of $w(\e,t)$. The computation is outlined in Appendix A and
the result \eq{des_modeenergy2} is:
\be \frac{\partial w(\e,t)}{\partial t}= \left( \frac{\e
w(\e,t)}{Q_2(t)}\right) \frac{\partial E}{\partial t}+\left(
\frac{K\Delta^2 g (\e)}{2}-\frac{a_c\e
w(\e,t)}{Q_2(t)}\right) A(t)~~. \label{des_modeenergy} \ee
In equilibrium one-time quantities are stationary. So the derivatives
of both the energy and $w(\e,t)$ are zero. From \eq{des_energy} and imposing
stationarity for the energy, we obtain the relation \eq{des_q2eq}
between the temperature and $Q_2^{\rm{eq}}$. From \eq{des_modeenergy}
and imposing that the derivative of the $w(\e,t)$ is zero, we obtain
the following relation:

\be \e w^{\rm{eq}}(\e) =\frac{g(\e)}{\widehat{\e}}Q_2^{eq}=
\frac{T}{2} g(\e)~~~~,\label{des_eqcond}\ee valid for any $\e$ in
equilibrium. Eq. \eq{des_eqcond} is the equipartition theorem
for each energy mode $\e$. It states that the average energy per
degree of freedom is equal to $\frac{T}{2}$.

Next we consider the mode-density relaxation at
$T=0$. Eq.~\eq{des_modeenergy} can be solved in the asymptotic long-time
limit using the results from the adiabatic approximation
\eqq{des_accept_adi}{des_energy_adi} by noting that $\frac{\partial
E}{\partial t} << A(t)$ and $\frac{\partial w(\e,t)}{\partial t} \to 0$ yielding:
\be \e w(\e,t)\simeq \frac{g(\e)}{\widehat{\e}}Q_2(t) ~~~~\forall~
\e~~~~,\label{des_equipart}\ee
which can also be written using the adiabatic
approximation \eq{des_adiabatic} as:
\be \e w(\e,t)\simeq g(\e)E(t)~~~\rightarrow~~~~~~~\frac{\partial
w(\e,t)}{\partial t}<0 ,~~~~{\rm for}~~ t \to
\infty~~~~.\label{eqfelix0}\ee
This last result, which is valid for all values of $\e$ in the
asymptotic (adiabatic) long time limit, is based on the fact that the
energy is a monotonic decreasing function of time. On the other hand,
for low enough values of $\e$ the first term in the rhs
of \eq{des_modeenergy} can be neglected and we get:
\be \frac{\partial w(\e,t)}{\partial t}=
\frac{K\Delta^2g(\e)}{2}A(t)> 0~~~~~~\e \to 0~~~~.\label{eqfelix1}\ee
We conclude that, for long-enough times and for continuous
distributions of $g(\eps)$ with support on $\e$-values extending down to
$\e=0$, there must exist a value $\epsilon^*$ such that $\frac{\partial
w(\e^*,t)}{\partial t}=0$, i.e.  $\frac{\partial w(\e,t)}{\partial t}$
changes sign at $\e=\e^*$. From \eqqq{des_modeenergy}{eqfelix0}{eqfelix1} we can distinguish
three different regions of $\e$ values depending on the sign of
$\frac{\partial w(\e,t)}{\partial t}$:
\be {\rm{Region~~I~:}}~~~~~~~~ \e
w(\e,t) > \frac{g(\e)}{\widehat{\e}} Q_2(t)
\left(\frac{1}{1-\frac{1}{A(t)a_c}\frac{\partial E}{\partial
t}}\right)~~~~\longrightarrow~~~~~\frac{\partial w(\e,t)}{\partial t} <
0~, \label{des_condition1} \ee
\be {\rm{Region~~II~:}}~~~~~~~~ \e w(\e,t) <
\frac{g(\e)}{\widehat{\e}} Q_2(t)
\left(\frac{1}{1-\frac{1}{A(t)a_c}\frac{\partial E}{\partial
t}}\right)~~~~\longrightarrow~~~~~\frac{\partial w(\e,t)}{\partial
t} > 0~,\label{des_condition2}  \ee
\be {\rm{Threshold ~~Boundary ~:}}~~~~~~~~ \e^* w(\e^*,t) =
\frac{g(\e^*)}{\widehat{\e}} Q_2(t)
\left(\frac{1}{1-\frac{1}{A(t)a_c}\frac{\partial E}{\partial
t}}\right)~~~~\longrightarrow~~~~~\frac{\partial
w(\e^*,t)}{\partial t} = 0~.\label{des_condition3} \ee
Equation \eq{des_condition3} defines the time-dependent threshold
$\e^*$. Modes above the threshold are described by \eq{des_condition1}
and have a negative time-derivative of the mode-energy density $w(\e,t)$
while modes below the threshold are described by \eq{des_condition2}
and have a positive time-derivative of the corresponding mode-energy
density. For distributions $g(\e)$ whose support of $\e$ values does
not extend down to zero it does not necessarily exist a value $\e^*$
which satisfies \eq{des_condition3}. In such cases the two
regions \eqq{des_condition1}{des_condition2} may still exist but not
the threshold boundary.

At zero temperature the system is always out-of-equilibrium: the
acceptance decays as $A(t)\approx \frac{1}{t\log t}$
\eq{des_accept_adi} and the energy decays as $E(t)\approx
\frac{1}{\log t}$ \eq{des_energy_adi}, (these results being valid at
leading order in $1/\log t$). Therefore in the long-time limit:
\be \frac{1}{A(t)a_c}\frac{\partial E}{\partial t}\simeq
-\frac{1}{\log t}~~\longrightarrow~~0~.\ee
The asymptotic time-dependence of the threshold \eq{des_condition3} is
then given by the simple expression:
\be \e^* w(\e^*,t) = \frac{g(\e^*)}{\widehat{\e}}
Q_2(t)~. \label{des_threshold}\ee
From the dynamical equations \eq{des_modeenergy}
we can check  that modes with $\e>\e^*$ verify the equipartition
theorem at all times \eq{des_equipart},
\be \e w(\e,t) = \frac{g(\e)}{\widehat{\e}} Q_2(t)~~~,~~~\rm{for}
~~~~ \e > \e^* \label{des_equipartition}\ee
so these modes are partially equilibrated at the temperature $\Theta
(t)$ \eq{des_teff_adi}, i.e they have equilibrated over the hyper
surface (in phase space) of constant energy $E(t)=\frac{\Theta
(t)}{2}$.

From \eq{des_adiabatic}, the value of $Q_2(t)$ decreases as the energy
does. For distributions $g(\e)$ with support on $\e$-values extending
down to $\e=0$, the threshold value $\e^*$ decays asymptotically to
zero as a function of time. The time decay depends in a non trivial
way on the form of the distribution $g(\eps)$ for $\eps\to 0$. For a
distribution with finite weight at zero $\e$ the decay seems to be a
power law (see below). Unfortunately we have been unable to find a
general analytical expression for such decay.

\begin{figure}[hbp!]
\begin{center}
\includegraphics*[width=10cm,height=8cm]{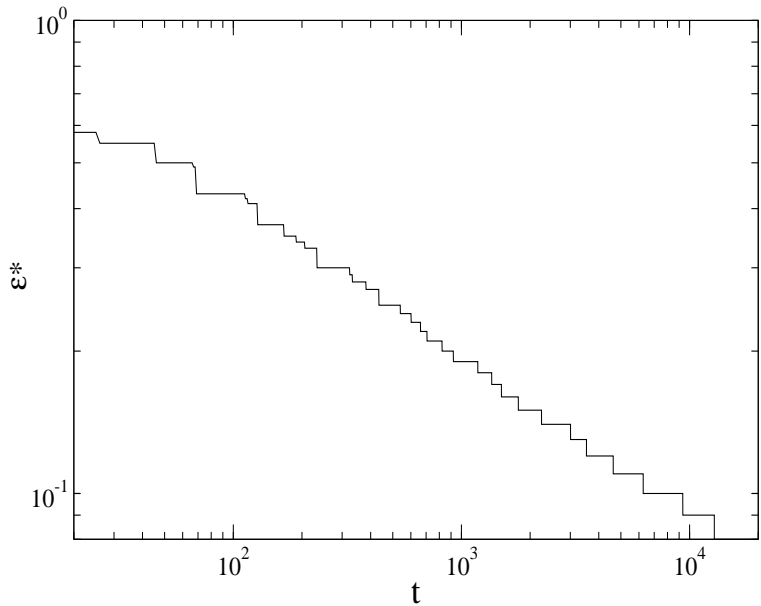}
\vskip 0.1in \caption{Time evolution (in Monte Carlo steps) of the
threshold energy value \eq{des_threshold} for a system of oscillators
with $g(\e)=1$, $\e\in [0,1]$ for $100$ spring constants uniformly
distributed between 0.01 and 1. As time goes on, $\e^*$ decreases
meaning that modes with longer time-scales progressively
equilibrate.\label{fig3}}
\end{center}
\end{figure}

By numerically solving \eq{des_threshold} we can compute the
time-evolution (in Monte Carlo steps) for the energy threshold. In
Fig.~\ref{fig3} we show the time evolution of $\e^*$ for a system that
consists of 100 types of oscillators with $g(\e)=1$ and spring
constants uniformly distributed in the range $\e=0.01-1$. It is clear
that, as time goes on, the threshold energy diminishes and more and
more oscillators partially equilibrate. As the system under
consideration has 100 different types of oscillators then the
solutions of \eq{des_threshold} are discretized also. This is the
reason why the time evolution for $\e^*$ shows a stepwise shape.
As the energy of the mode gives the characteristic time-scale of the
corresponding oscillators (see Sec.\ref{subsec2} below and Appendix B,
in particular \eq{des_eqtime}), those oscillators which have a
higher value of the spring constant $K\e$ also have a shorter
relaxational time-scale. As time goes on, the modes with $\e>\e^*$ are
the first to partially equilibrate, while those modes with lower $\e$
equilibrate over longer time scales as they are cached up by the
decreasing threshold value $\e^*$ \eq{des_threshold}.

\begin{figure}[hbp!]
\begin{center}
\includegraphics*[width=10cm,height=8cm]{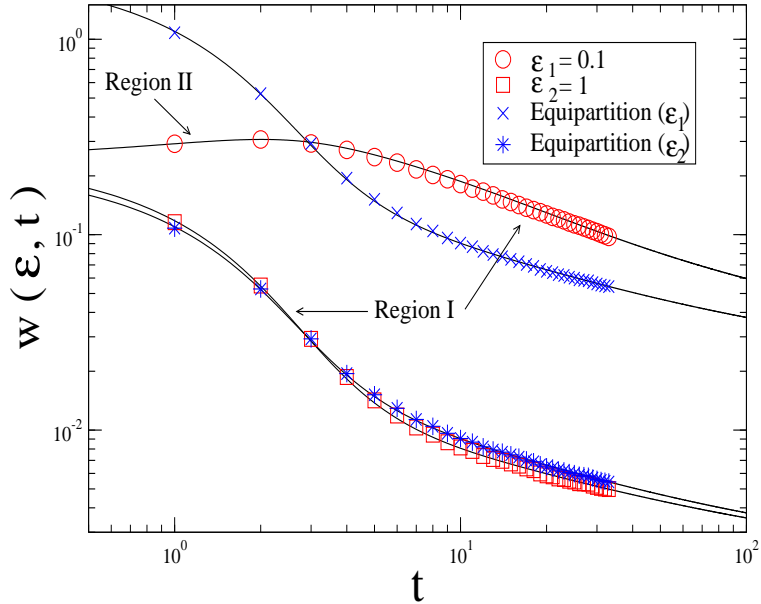}
\vskip 0.1in \caption{(Color online) Zero-temperature time evolution
(measured in Monte Carlo steps) of $w(\epsilon,t)$ \eq{des_modeenergy}
for a system with a binary distribution of the disorder
$g(\e)=(1/2)(\delta(\e-\e_1)+\delta(\e-\e_2))$ with $\e_1=1$ and
$\e_2=0.1$. We also plot $w(\epsilon,t)$ coming from the corresponding
equipartition relations \eq{des_equipartition} for the two modes of
the system. The two continuous curves in the top correspond to $\e_1$
while the two curves at the bottom correspond to $\e_2$. The symbols
correspond to the Monte Carlo simulation for a system with 5000
oscillators in each mode. The continuous lines correspond to the
numerical solution of \eq{des_modeenergy} and \eq{des_equipartition}
corresponding to the exact and the adiabatic solution
respectively. Different regions I \eq{des_condition1} and II
\eq{des_condition2} according to the sign of the time derivative of
$w(\epsilon,t)$ are also shown.\label{fig4}}
\end{center}
\end{figure}
A summary of this behavior is shown in Fig.~\ref{fig4} for the case of a
binary distribution of the disorder
$g(\e)=(1/2)(\delta(\e-\e_1)+\delta(\e-\e_2))$. For this particular
example we note that equation \eq{des_threshold} is only defined for
$\e=\e_1,\e_2$ so it does not have a well defined threshold solution
$\e^*$.  However, as we have already mentioned earlier, we can
distinguish the two regions \eq{des_condition1} and \eq{des_condition2}
corresponding to different signs of
$\frac{\partial w(\e,t)}{\partial t}$. This is shown in Fig.\ref{fig4}
where we show the time evolution of $ w(\e,t)$ at zero
temperature \eq{des_modeenergy}. We also plot the solution for $w(\e,t)$
using the equipartition theorem \eq{des_equipartition} for each of the
two modes. We can clearly see that the oscillators with the largest
value of $\e$ partially equilibrate very fast to the temperature $\Theta
(t)$ \eq{des_teff_adi} (the two curves at the bottom), while the two upper
curves deviate from each other even at long times. In Fig.\ref{fig4} we
also plot the results obtained from a Monte Carlo simulation (symbols)
and compare them with the numerical solution of the exact
equation \eq{des_modeenergy} and the adiabatic
approximation \eq{des_equipartition} (both in lines). The adiabatic
solution reproduces well the exact solution at large values of $\eps$
(region I) whereas it deviates from it at the lower values of $\e$
(region II). The collapse between Monte Carlo simulations (symbols) and
the numerical solutions (lines) is excellent. 

To sum up, in the non-equilibrium regime we have found two different
classes of oscillators in the system: on the one hand there are
partially equilibrated oscillators ($\e > \e^*$) for which the
equipartition theorem \eq{des_equipartition} holds out of equilibrium
as in the model without disorder \cite{des_bpr}; on the other hand,
there are non-equilibrated modes ($\e < \e^*$) for which the
equipartition theorem does not hold. This fact has been illustrated
with the binary case of a system with two different values of $\eps$,
i.e two different time-scales. 

In the next section we will define an effective temperature for
each energy mode (each time-scale) in the system. Again we will
distinguish between the partially equilibrated modes and
the rest.
\subsection{MODE CORRELATIONS AND RESPONSES}
\label{subsec2}

We can define the correlation and the response functions for each energy
mode. Let us define the mode-correlation $C(\e,t,s)$ as:
\be C(\e,t,s) = <\frac{1}{N} \sum_i \delta (\e_i - \e) x_i(s)x_i(t)>~~,
\label{mode_correl}
\ee 
which is the correlation between the positions of the oscillators
which have the same spring constant $K\e$. In this case, similar
calculations as in \cite{des_bpr}, show that the dynamical evolution of
the mode-correlation function is given by:
\be \frac{\partial C(\e,t,s)}{\partial t}= \left( \frac{\partial
E}{\partial t}-a_cA(t) \right)
\frac{\e}{2Q_2}C(\e,t,s)~~. \label{des_corrvst}\ee
We can compute the dynamic evolution for the mode-response after
perturbing the Hamiltonian:
\be H_{\e}=H_0-hM(\e)~~~~, \ee
where $H_0$ is defined in \eq{des_hamilt} and the perturbation is
coupled to the magnetization of that particular mode of energy
$\epsilon$. We define the mode-magnetization as:
\be M(\e,t)=<\frac{1}{N} \sum_i
\delta(\e_i-\e) x_i(t)>~~.\label{des_modemag}\ee
The mode-response function $G(\e,t,s)$ is defined as usual,

\be G(\e,t,s)= \left(\frac{\delta M(\e,t)}{\delta
h(s)}\right)_{h(s)\rightarrow 0}~~~~t>s~~~~. \label{mode_response}\ee

Computing the dynamical equation for the mode-magnetization and
deriving it respect to the external field $h$, we obtain the dynamical
equation for the mode-response function:
 \beq \frac{\partial
G(\e,t,s)}{\partial t}= \left( \frac{\partial E}{\partial
t}-a_cA(t) \right) \frac{\e}{2Q_2}G(\e,t,s)+\left( \frac{\partial
E}{\partial t}-a_cA(t)
\right)\frac{g(\e)}{2KQ_2}\delta(t-s).\label{eqfelix2}
\eeq
Note that the evolution of the mode-correlation and the
mode-response are identical and differ only from a delta function
term. Equations \eqq{des_corrvst}{eqfelix2} can be solved exactly and
allow us to make a deeper analysis of the non-equilibrium behavior of
the system. The equilibrium behavior of these functions is analyzed in
the Appendix B.

In the general nonequilibrium regime the correlation function and the
response function can be written \eqq{des_corrgen}{des_respgen}:
\be C(\e,t,s)=\frac{2}{K}w(\e,s)\exp\left[ -(t-s)/\tau
(\e,t,s)\right]~~~~t>s,\label{des_corrtau}\ee
\be G(\e,t,s)=\frac{F(\e,s)g(\e)}{\e
K}\exp\left[ -(t-s)/\tau
(\e,t,s)\right]~~~~t>s,\label{des_resptau}\ee
where $F(\e,s)$ is given in \eq{des_corrf} and we have defined a
time-dependent relaxation time $\tau(\e,t,s)$ \eq{des_tdtime2}:
\be \tau(\e,t,s)=
\frac{(t-s)}{\e}\left(\frac{1}{\int_s^t\frac{f(x)}{2Q_2(x)}dx}
\right)~~~~t>s.\label{des_tdtime}\ee
The relaxation time of a given oscillator is inversely proportional to
the value of its spring constant $\e$. This result indicates that each
type of oscillator (each $\e$) can be considered as a single variable
(evolving according Langevin dynamics) with a time-dependent
relaxation time $\tau(\e,t,s)$ given by the
expression \eq{des_tdtime}. Note that the distribution of disorder
fully determines correlations and responses through the functions
$f(x)$ and $Q_2(x)$.

The full correlation function is then given by,
\be C(t,s)=<\frac{1}{N} \sum_i x_i(s)x_i(t)>=\int_0^\infty d\e
C(\e,t,s)~~~~,\label{full_corr}\ee
(an analogous expression is valid for the
response function). The relaxation of the full correlation function depends strongly on
the disorder distribution $g(\e)$. In equilibrium, the full
correlation function can be written
using \eqqq{des_correqap}{des_eqtime}{des_gamma},
\be C^{\rm eq}(t-s)=\int_0^\infty \frac{T}{K}\frac{g(\e)}{\e}\exp \left( -\e
\left(\frac{(t-s)K}{\gamma} \right)
\right)d\e~~~~.\label{des_fulleqcorr}\ee
Defining the parameter $\tau=\left(\frac{(t-s)K}{\gamma} \right)$ this
correlation \eq{des_fulleqcorr} is given by the following Laplace transform,
\be C^{\rm eq}(\tau)=\int_0^\infty h(\e)\exp \left(-\e\tau
\right)d\e~~~~,\label{ctau}\ee where we have defined $h(\e)$ as: \be
h(\e)=\frac{T}{K}\frac{g(\e)}{\e}~~~~.\ee
The equilibrium correlation function  $C^{\rm eq}(\tau)$ is well defined
whenever the integral \eq{ctau} converges, i.e. for $g(\e\to 0)\sim\e^{a}$
with $a>0$. In this case, also the entropy constant in \eq{free} is
finite.  When the integral \eq{ctau} does not converge it means that the
largest contribution to the full correlation function is determined
by diffusive behavior of the $\e\to 0$ modes.

A relevant feature of the DOM is that it can describe the
heterogeneous behavior observed in glass formers (see the final
discussion in Section~\ref{sec7}). The decorrelation time
in the DOM can be defined in terms of the integrated relaxation time
\cite{des_felixsoll}:
\be \tau_{\rm relax}=\frac{\int_0^{\infty}C^{\rm
eq}(\tau)d\tau}{C^{\rm eq}(0)}=\frac{\widehat{\e_{-2}}}
{\widehat{\e_{-1}}}\frac{\gamma}{K}=\sqrt{\frac{T\widehat{\e}\pi}{2\Delta^2K}}\frac{\widehat{\e_{-2}}}
{\widehat{\e_{-1}}}\exp \left[ \frac{K\Delta^2
\widehat{\e}}{8}~~\frac{1}{T} \right]
\label{taurelax}
\ee
where we have used \eq{moments} (with $\widehat{\e}=\widehat{\e_1}$)
and the expression for the viscosity $\gamma$ \eq{des_gamma2}. This
expression shows that the relaxation time is thermally activated and
that the main effect of the disorder on the relaxation time is through
the activation barrier which is proportional to $\widehat{\e}$. The
other moments of the disorder distribution
${\widehat{\e_{-1}}},{\widehat{\e_{-2}}}$ only affect the prefactor of
the relaxation time.

In general, the distribution of disorder $g(\e)$ can be chosen in order to
reproduce any given relaxation function by computing the inverse Laplace
transform of the full correlation $C^{\rm eq}(\tau)$. As an example we
consider a few cases:

\begin{itemize}
\item{Binary distribution.} Consider a binary system with one half of the
oscillators having $\e=\e_1$ and the other half having $\e=\e_2$ where $\e_2 >
\e_1$. In this case, the full correlation
function in equilibrium \eq{des_fulleqcorr} is simply the sum of two
exponentials,
\be C^{\rm eq}(t-s)=\frac{T}{2K\e_1\e_2} \left(\e_2 \exp
\left(-\frac{K\e_1}{\gamma}(t-s) \right)+\e_1 \exp
\left(-\frac{K\e_2}{\gamma}(t-s) \right) \right)~~~~.\label{corr_bin}\ee
For long times, the full correlation \eq{corr_bin} is dominated by the
first exponential. For a general uniform discrete distribution with $n$
finite different values of $\e$, the relaxation of the full correlation will
be the sum of $n$ exponentials of the type \eq{des_correqap}.

\item{Continuous distribution.} Consider now the continuous distribution
$g(\e)=A\e^\alpha \exp(-\e)$ where $A$ is the normalization factor. In
this case, the full correlation function in equilibrium can be computed
from \eqq{full_corr}{des_fulleqcorr} giving a power law decay:
\be C^{\rm eq}(t-s)=\frac{T}{K\alpha}\left(\frac{1}{1+\frac{K}{\gamma}(t-s)}
\right)^\alpha~~~~.\ee

\item{Stretched exponential relaxation.} Let us consider now the
inverse problem of finding the disorder distribution for a correlation
of the type:
\be C^{\rm eq}(t-s) \propto \exp\left(-A\sqrt{t-s}\right)~~~~.\ee
Taking the inverse Laplace transform, we get for the disorder
distribution:
\be g(\e)\propto
\e^{-\frac{1}{2}}e^{-\frac{A^2\gamma}{4K\e}}~~~~.\label{des_stret}\ee
where $A$ is a positive constant. Note that this distribution is
non-normalizable since it decays too slowly in the $\e\to\infty$
limit. This is not a problem since the large $\e$ limit only
influences the short $(t-s)$ behavior of the correlation
function. Therefore by introducing a cutoff for $g(\e)$ in the large
$\e$ limit we do not modify the stretching of the relaxation function.
For the large-$t$ generic stretching exponential behavior $C^{\rm
eq}(\tau) \sim \exp\left(-A\tau^{b}\right)$ we have, in the $\eps\to
0$ limit, $g(\e)\sim \exp(-B/\e^{\frac{b}{1-b}})$ with $B$ a positive
constant. Taking $b=1/2$ we recover the result (\ref{des_stret}) in
the $\eps\to 0$ limit. The exponential relaxation corresponds to the
case $b\to 1$ and the $g(\e)$ has a gap. In general, the large $t$
behavior of $C^{\rm eq}(\tau)$ is determined by the $\e\to 0$
dependence of $g(\e)$.

\end{itemize}

From this analysis we can see that the DOM is a very general model.
The distribution of the disorder can be chosen in order to fit any
experimental data just computing the inverse Laplace transform of the
correlation.

\subsection{MODE EFFECTIVE TEMPERATURES}
\label{subsec3}
Building up on the ideas by Cugliandolo, Kurchan and Peliti
\cite{des_kurchan} we can define an effective temperature for each mode
and analyze how the different degrees of freedom in the system
relax. From \eq{des_teff} we can define an
effective temperature for each mode:
\be T_{\textrm{eff}}(\e,s) =\left(\frac{\frac{\partial
C(\e,t,s)}{\partial s}}{G(\e,t,s)}
\right)=\frac{2\epsilon}{g(\e)}\left( w(\e,s) +
\frac{1}{F(\e,s)}\frac{\partial w(\e,s)}{\partial
s}\right)~~.\label{teffdes}\ee
The mode effective temperature $T_{\textrm{eff}}(\e,s)$ only depends
on the lowest or waiting time $s$ but not on the largest time
$t$. This is a typical feature of mean-field structural glass models~\cite{des_felixcris}.

As mentioned in Sec~\ref{intro}, we can also define an
effective temperature in the Fourier space \eq{des_teffF} for each
mode:
\be T_{\textrm{eff}}^F(\eps,\Omega,s)=\Omega
\frac{S(\eps,\Omega,s)}{\widehat{\chi}''(\eps,\Omega,s)}
~~~~,\label{eqfelix3}\ee
where $S(\eps,\Omega,s)$ and $\widehat{\chi}(\eps,\Omega,s)$ are
defined as \cite{des_felixsoll}:
\be S(\eps,\Omega,s)=\int_s^\infty dt \cos (\Omega t) C(\e,t,s)~~~~,
\ee \be \widehat{\chi}(\eps,\Omega,s)=i\Omega \int_s^\infty dt
\chi (\eps,t,s)\exp{(-i\Omega t)}~~~~,
\label{eqfelix4}
\ee
In equilibrium the power spectrum can be written as: \be S^{\rm
eq}(\eps,\Omega)=\int_0^\infty d\tau \cos (\Omega \tau) C^{\rm
eq}(\e,t-s=\tau)~~~~, \ee where $C^{\rm eq}(\e,t,s)$ is given
in \eq{des_correqap}. The result is a Lorentzian function:
\be S^{\rm eq}(\eps,\Omega)= \frac{T}{\gamma}
g(\e)\left(\frac{1}{\Omega^2+\left( \frac{\e^2
K^2}{\gamma^2}\right)}\right)~~~~.\ee
The full power spectrum is expressed in terms of the full
correlation function \eq{full_corr} and is given by the integral of the
Lorentzian weighed by $g(\e)$:
\be S^{\rm eq}(\Omega)=\int_0^\infty d\tau \cos (\Omega \tau) C^{\rm
eq}(\tau)=\frac{T}{\gamma}\int_0^\infty
g(\e)\left(\frac{1}{\Omega^2+\left( \frac{\e^2
K^2}{\gamma^2}\right)}\right)d\e ~~~~.\label{des_pseq}\ee
As for the full correlation function the disorder distribution
$g(\e)$ can be chosen to fit the experimental power spectrum data by
using the relation \eq{des_pseq}. In \eq{eqfelix4} we have used the
susceptibility, which in terms of the response is given by,
\be
\chi(\e,t,s)= \int_s^t G(\e,t,\tau) d\tau ~~~~.\ee
Note that there is a factor 2 missing in \eq{eqfelix3} as compared
to the standard definition of the FDT in frequency
space \cite{des_felixsoll}. This is due to the fact that we are
assuming that the integral \eq{eqfelix4} extends from $s$ (when the
measure of the relaxation function starts) to $\infty$ rather than
from $-\infty$ to $\infty$ as in the equilibrium case. The calculations
are shown in Appendix D. With these
definitions we obtain for the effective temperature in Fourier
space \eq{des_fourierteff}:
\be T_{\textrm{eff}}^F(\eps,\Omega,s)\equiv T_{\textrm{eff}}^F(\eps,s)=
\frac{2\e}{g(\e)}w(\e,s)~~~~.\label{des_teffF2}\ee
Interestingly, the effective temperature in Fourier space does not
depend on the frequency and only depends on the waiting time $s$ and
the mode $\eps$ considered. We can observe that the result in Fourier
space \eq{des_teffF2} is equal to the first term of \eq{teffdes} in the
time domain so the subleading corrections introduced by the second
term in \eq{teffdes} are absent in $T_{\textrm{eff}}^F(\eps,s)$.

\begin{figure}[hbp!]
\begin{center}
\includegraphics*[width=10cm,height=8cm]{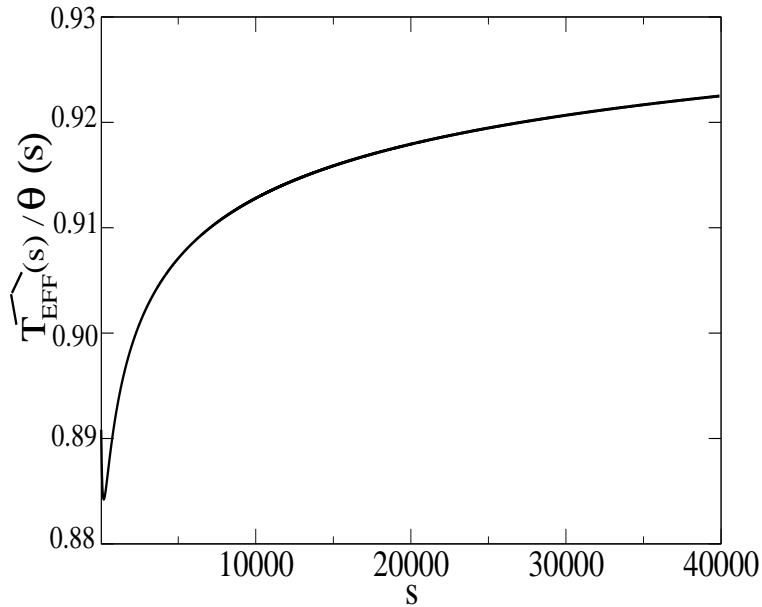}
\vskip 0.1in \caption{Time evolution (measured in Monte Carlo steps)
of the full effective temperature (normalized by its asymptotic value
\eq{des_teff_adi}) for the same system considered in Fig.\ref{fig3}.
\label{fig4b}}
\end{center}
\end{figure}

Let us note that it is also possible to define a ``full'' effective
temperature $ \widehat{T_{\textrm{eff}}}(s)$ (either in the time or
frequency domains) from the full correlation function \eq{full_corr}
and the analogous expression for the response function. From the
analysis carried out in Section~\ref{subsec1} we have that $\e^*\to 0$
asymptotically. The full effective temperature is then determined by
the contribution to the full correlation and full response from the
modes in region I, $\e > \e^*$. In Fig.\ref{fig4b} we can see the
time-evolution (in Monte Carlo sweeps) of the normalized full effective
temperature in time domain for the system considered in
Fig.\ref{fig3}, with 100 types of oscillators with spring constants
uniformly distributed in the range $\e=0.01-1$. As time goes on, more
and more oscillators partially equilibrate and the full effective
temperature approaches asymptotically to $\Theta (s)$
\eq{des_teff_adi}.

\begin{figure}[hbp!]
\begin{center}
\includegraphics*[width=10cm,height=8cm]{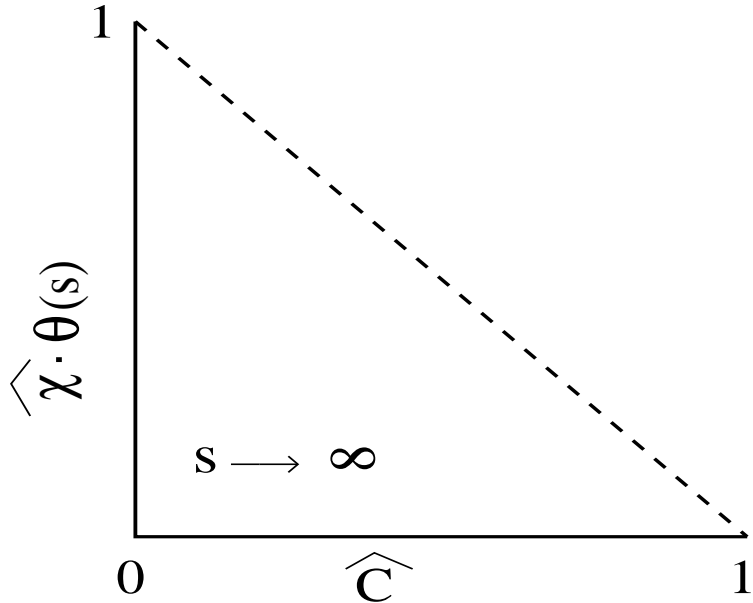}
\vskip 0.1in \caption{Asymptotic FD plot: the full integrated response
function normalized by the asymptotic effective temperature
\eq{des_teff_adi} as a function of the normalized full correlation
\eq{prediction} for $s$ fixed.
\label{figfdt}}
\end{center}
\end{figure}

In the glass literature it is sometimes useful to construct the
so-called fluctuation-dissipation (FD)
plots~\cite{des_sollmay}. Generally, a FD plot is a plot of the
integrated-response versus the correlation function. In the DOM, the
full correlation \eq{full_corr} at equal times $C(s,s)$ depends on
$s$. Therefore, for a meaningful FD plot we should consider the
normalized full correlation $\widehat{C}(t,s)=C(t,s)/C(s,s)$ and the
full integrated-response given by:
\be \widehat{\chi}(t,s)=\int_s^t d\tau \int_0^\infty d\e
G(\e,t,\tau)~~~~.\label{des_fullresp}\ee
The FD plot consists on representing $\widehat{\chi}$ as a function of
$\widehat{C}$ for a fixed waiting time $s$ and varying $t$. At
equilibrium this plot gives a straight line with slope $-1/T$ where T is
the temperature of the bath. In the DOM, for asymptotic waiting times
($s\to\infty$) we find, for a fixed $s$,
\be \widehat{\chi}(\widehat{C})= \frac{1}{\Theta(s)} \left(
1-\widehat{C}\right)~~~~.\label{prediction}\ee
Therefore, for a given asymptotic waiting time $s$ the FD plot gives a
straight line with negative slope equal to the inverse of the
asymptotic effective temperature \eq{des_teff_adi}. In
Fig.\ref{figfdt} we represent the resulting asymptotic FD plot which
has also been obtained before in the OM model~\cite{des_felixnoeq2} as
a particular case. Straight FD plots are characteristic of the
one-step behavior of structural glasses. In Fig.\ref{figfdt} we can
also see the absence of an equilibrium relaxation (usually called
$\beta$-relaxation) for small responses. This unrealistic feature is
due to the mean-field nature of the model.

Next we consider in detail the effective temperature for the different
types of modes. As we will see all modes in region I give the same
mode effective temperature value which is equal to the value of the
full effective temperature.

\subsubsection{Region I of partially equilibrated modes: $\e > \e^*$}

For the partially equilibrated modes, i.e $\epsilon > \epsilon^*$,
we just keep the first term in \eq{teffdes} (the derivative of
the $w(\e,s)$ goes to zero faster than $w(\e,s)$ itself). Therefore,

\be T_{\textrm{eff}}(\e,s)\simeq
\frac{2\epsilon}{g(\e)}w(\e,s)~~~~{\rm{with}} ~~~~ \e w(\e,s) =
\frac{g(\e)}{\widehat{\e}} Q_2(s)~, \ee and the effective temperature
reads: \be T_{\textrm{eff}}(\e,s)\simeq\frac{2}{\widehat{\epsilon}}
Q_2(s)=\Theta(s)~~,\label{des_infty} \ee which coincides with the
expression derived within the adiabatic approximation
\eq{des_teff_adi} (up to subleading logarithmic corrections of order
$1/\log(t)$).

For the effective temperature in Fourier space we get the same result
but without logarithmic corrections:
\be T^F_{\textrm{eff}}(\e,s)= \frac{2\epsilon}{g(\e)}w(\e,s)=
\Theta(s)~~.\ee
All modes which are partially equilibrated share the same effective
temperature that we define as $ \Theta(s)$. As the time $s$ goes on,
more and more oscillators achieve this value $ \Theta(s)$ and, in the
asymptotic limit, all oscillators will thermalize to this effective
temperature.

\begin{figure}[hbp!]
\begin{center}
\includegraphics*[width=10cm,height=8cm]{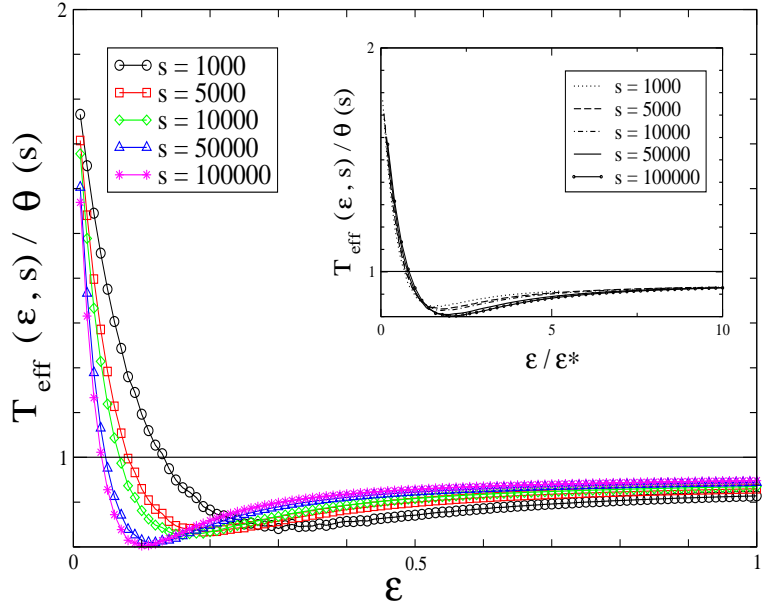}
\vskip 0.1in \caption{(Color online) The dependence of the effective
temperature (in time domain) as a function of the value of $\e$ for
the same system as in Fig.\ref{fig3} with 100 types of oscillators
with $\e$-values equally spaced and uniformly distributed between 0.01
and 1.00. Data is shown for different waiting times at zero
temperature. In the inset we show the scaling of the effective
temperature with the threshold energy $\e^*$.\label{fig5}}
\end{center}
\end{figure}

In Fig.\ref{fig5} we can see the dependence of the effective
temperature \eq{teffdes} on the energy mode at a given time.  We plot
the effective temperature normalized by $\Theta(s)$ for different
values of $s$ at zero temperature.  As we increase the waiting time
$s$ those oscillators with higher value of $\e$ (shorter relaxation
time according to \eq{des_eqtime}) tend to achieve an effective
temperature $T_{\rm{eff}}(\e,s)$ equal to $\Theta(s)$. Whereas those
oscillators with lower values of $\e$ have an effective temperature
greater than $\Theta(s)$. From the figure we note that the ratio
$T_{\rm{eff}}(\e,s)/\Theta(s)$ for $\e>\e^*$ is smaller than 1 due to
the logarithmic correction coming from the second term in the r.h.s of
\eq{teffdes} whereas such corrections are not present in
$T_{\textrm{eff}}^F(\eps,s)$. Therefore, as the waiting time
increases, lower energy modes progressively reach partial
equilibrium. In the inset of Fig.\ref{fig5} we plot the normalized
effective temperature as a function of $\e/\e^*$ showing good
collapse. The shape of Fig.\ref{fig5} is expected to be quite
general. The same qualitative picture was found for a Gaussian
distribution for the disorder or an exponential one (data not shown).

\begin{figure}[hbp!]
\begin{center}
\includegraphics*[width=10cm,height=8cm]{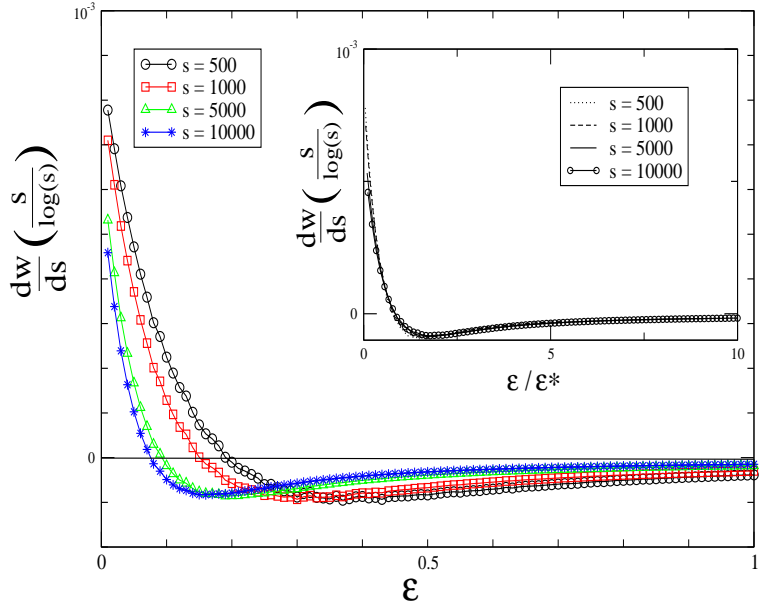}
\vskip 0.1in \caption{(Color online) The derivative of the mode-energy density
multiplied by $\frac{s}{log(s)}$ as a function of $\e$ for the
same system as in Fig.\ref{fig5} In the inset we show the same plot as a
function of $\e/\e^*$.\label{fig6}}
\end{center}
\end{figure}

Modes around the threshold, $\e \simeq \e^*$, are not yet partially
equilibrated. By plotting $\frac{\partial w(\e,s)}{\partial s}$ as a
function of $\e$ we can identify the width $\Delta \e$ of the region 
separating the partially equilibrated modes from the rest. This is
shown in Fig.\ref{fig6} for the same system considered in
Fig.\ref{fig5}. There, we have multiplied $\frac{\partial
w(\e,s)}{\partial s}$ by a factor $\frac{s}{\log(s)}$ which is the
inverse of the leading order of the decay of $\frac{\partial
w(\e,s)}{\partial s}$ computed in the adiabatic approximation. In the
inset we can clearly see that this quantity again scales as a function
of $\frac{\e}{\e^*}$. From such scaling we note that the width $\Delta
\e$ around the threshold scales as $\e^*$ and therefore, as $\e^*$
decays to 0, the region around the threshold becomes progressively
thinner relative to the support of $g(\e)$. This allows us to
define the threshold boundary as the asymptotically thin sector of
$\e$-values that separates region I from region II.

\subsubsection{Region II of diffusive modes: $\e < \e^*$}

Using \eq{des_modeenergy}, the expression for the effective
temperature \eq{teffdes} can also be written as:

\be T_{\textrm{eff}}(\e,s)= 2\Theta(s)- \frac{2\e
w(\e,s)}{g(\e)}~~~.\label{des_teffinf}\ee
As a consequence those degrees of
freedom with larger time-scales are \textit{hotter} than partially
equilibrated ones. This result can be checked in Fig.\ref{fig5} where the
oscillators in the region $\e<\e^*$ have an effective temperature higher
that $\Theta(s)$.

For finite times, in the limit $\e \to 0 $, the second term in the rhs
of \eq{des_teffinf} vanishes 
\be T_{\textrm{eff}}(\e \to 0)=2\Theta(s)~~~~,
\ee 
and the effective temperature is just twice the asymptotic effective
temperature (the effective temperature for the partially equilibrated
modes). Note that this value corresponds to the effective temperature of
a diffusive process \cite{des_parisi}, whose value is twice the
temperature as given by the Einstein relation $D=2T$ where $D$ is the
diffusion constant associated to the diffusive
variable $x$, $(\Delta x^2(t))\sim Dt$. This result can be intuitively
understood. Modes with $\e$ small do not contribute much to the total
energy \eq{des_hamilt}, so their motion is completely
uncorrelated from the rest of modes and become diffusive.

\begin{figure}[hbp!]
\begin{center}
\includegraphics*[width=10cm,height=8cm]{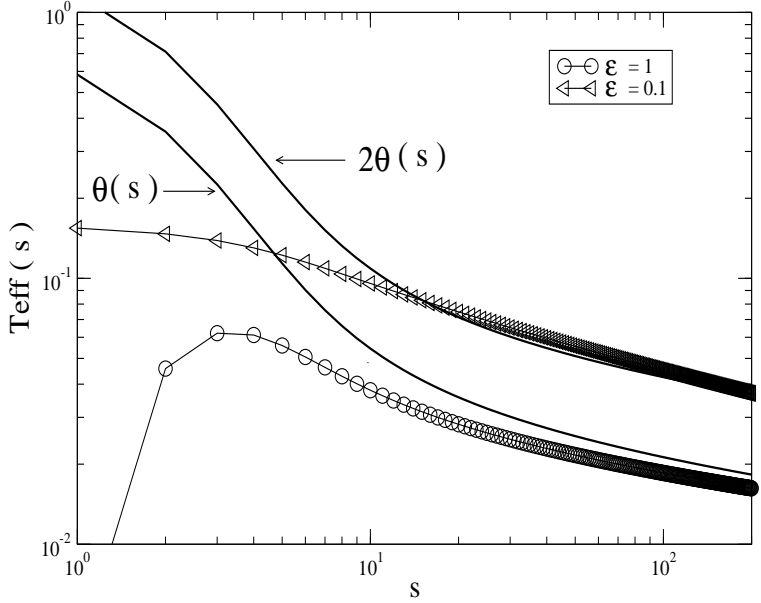}
\vskip 0.1in \caption{We plot the effective temperature for a binary
system: 20\% of oscillators have $\e=0.1$ (triangles) while the rest
have $\e=1$ (circles). We also plot the time-evolution, measured in
Monte Carlo steps, of $\Theta(s)$ and $2\Theta(s)$ \eq{des_infty} (full
lines) for the two modes.\label{fig7}}
\end{center}
\end{figure}

 In order to illustrate this result we show in Fig.\ref{fig7} the
effective temperature for a binary system. In this case, 20\% of the
oscillators have $\e=0.1$ while 80\% have $\e=1$. We can see how the
oscillators with larger $\e$ are partially equilibrated to the
effective temperature $\Theta(s)$ \eq{des_infty} while the oscillators
with smaller $\e$ have an effective temperature approximately equal to
$2\Theta(s)$ as given by \eq{des_teffinf}.

\section{FLUCTUATION RELATIONS AND THE EFFECTIVE TEMPERATURE}
\label{sec5}

The effective temperatures for the partially equilibrated modes can be
obtained by using a recently proposed fluctuation relation
\cite{des_crisfel}.  In \cite{des_felixosc} it has been suggested that
effective temperatures can be derived by assuming microscopic
reversibility between surfaces of constant energy. This relation has
been also proposed in the framework of mean-field spin glass models
\cite{des_seme} and applied to generalized oscillator models
\cite{des_felixnoeq2}. The effective temperature $\Theta(s)$ can be
computed from the ratio between the probabilities of having a change
in energy equal to $\delta E$ and $(-\delta E)$.  From
\eq{des_pdeltae},
\be \frac{P(\delta E)}{P(-\delta E)}=\exp\left(\delta
E\frac{\widehat{\epsilon}}{2Q_2 (s)}\right)=\exp\left(\frac{\delta
E}{\Theta (s)}\right)~~ \label{des_tefffluct}\ee
where we have used \eq{des_pdeltae}. This relation between forward and
reverse transitions only holds at the level of energies rather than
configurations. It should not be confused with the detailed balance
or reversibility property \eq{des_metro} of microscopic dynamics
which obviously holds at the level of configurations in and out of
equilibrium \eq{des_metro}. Equation \eq{des_tefffluct} defines a
nonequilibrium temperature that is equal to the effective temperature
obtained from the fluctuation-dissipation relation for the partially
equilibrated modes \eqq{des_teffF2}{des_infty}.

We can also extend this argument and compute the probability
$P_s(\emph{Q}_{\e})$ of having an instantaneous change in the energy $\emph{Q}_{\e}$ of
a given mode $\e$ at a given time $s$,
\be \emph{Q}_\e=<\e \frac{K}{2N}\sum_i \delta (\e_i-\e)x_i^2>= \e
w(\e,s)~~~~.\ee
The probability $P_s(\delta \emph{Q}_\e)$ is computed
in \eq{eqfelix5}. It is given
by:
\be P_s(\delta \emph{Q}_{\e})=\frac{1}{\sqrt{4\pi K\Delta^2 \e^2
w(\e,s)}} \exp \left[ -\frac{(\delta \emph{Q}_\e -
\frac{K\Delta^2}{2}\e g(\e))^2}{ 4K\Delta^2\e w(\e,s)} \right]
~~~~,\ee
and the fluctuation relation for each mode is given by:
\be \frac{P_s(\delta \emph{Q}_{\e})}{P_s(-\delta \emph{Q}_{\e})} =
\exp \left[ \delta \emph{Q}_\e \frac{g(\e)}{2\e
w(\e,s)}\right]=\exp\left[ \frac{\delta
\emph{Q}_\e}{T_{\textrm{eff}}^F(\eps,\Omega,s)}\right]~~~~.\label{ft1}\ee
Only for partially equilibrated modes, we can use \eq{des_infty} and
get,
\be \frac{P_s(\delta \emph{Q}_\e)}{P_s(-\delta \emph{Q}_\e)}=\exp
\left[ \frac{\delta \emph{Q}_\e}{\Theta(s)}\right]~~~~.\ee
This relation emphasizes the fact that only for the partially
equilibrated modes a unique effective temperature can be properly
defined. Otherwise, for non-equilibrated modes $\frac{2\e
w(\e,s)}{g(\e)} \neq \Theta(s)$

Recent work \cite{des_felixnoeq2,des_felixosc} has shown that this
effective temperature can also be obtained from microcanonical
relations relating observable changes. In the present case, we
consider the mode-magnetization \eq{des_modemag}. The probability of
having a change $\delta M_\e$ in the mode-magnetization is computed in
the Appendix C \eq{des_probm}. The rate of having a positive and a
negative change gives:

\be \frac{W^h(\delta M_\e)}{W^h(-\delta M_\e)}= \frac{W^0(\delta
M_\e)}{W^0(-\delta M_\e)} \exp\left( \delta M_\e
h\frac{\widehat{\e}}{2Q_2(s)} \right)= \frac{W^0(\delta
M_\e)}{W^0(-\delta M_\e)} \exp\left( \frac{\delta M_\e
h}{\Theta(s)} \right)~~~.\label{des_teffmicro}\ee

At difference with \eq{ft1} now the effective temperature $\Theta(s)$
describes mode-magnetization fluctuations for all values of $\eps$. This
is not unexpected due to the linear dependence of the observable $M_\e$
on the configurational variables \eq{des_modemag} and the Gaussian
character of the distribution \eq{eqc1} which determine that
\eq{des_probm} depends on $Q_2$ (i.e. on $\Theta(s)$) for all values of
$\eps$. Although not proven we suspect that this general dependence
might not hold in general, e.g. for generic non harmonic potentials~\cite{des_felixnoeq2}.

\section{SUMMARY OF RESULTS}
\label{sec6}

In this paper we have introduced an exactly solvable disordered model
for glassy dynamics, the disordered oscillator model (DOM). The model is
the disordered version of the oscillator model (OM) previously
introduced by one of us \cite{des_bpr} as a simple model with entropic
barriers for aging dynamics.  The DOM consists of a collection of
non-interacting oscillators or modes with different spring
constants. The model has trivial statics but is kinetically constrained and
has cooperative dynamics displaying glassy behavior at low temperatures. In
the DOM the relaxation time associated to each mode is inversely
proportional to the effective spring constant. This allows us to study
the relaxational properties (both in equilibrium and in the aging
regime) associated to each mode and the value of the effective
temperature \eq{des_teff} for the different degrees of freedom.

Dynamics in the DOM can be solved writing down a hierarchy of coupled
first-order differential equations~\eq{des_jer2}. These set of equations
can be closed by introducing a generating function and using the method
of characteristics or the adiabatic approximation at low
temperatures. The latter approach is much simpler and allows us to
define an effective temperature $\Theta(s)$ \eq{des_teff_adi} in the aging regime ($s$ being
the waiting time). The relaxation of one-time quantities appears similar
to the OM, just replacing the spring constant in the OM, $K$, by the average
spring constant $K\widehat{\e}$ in the DOM. 

\begin{figure}[hbp!]
\begin{center}
\includegraphics*[width=8cm,height=6cm]{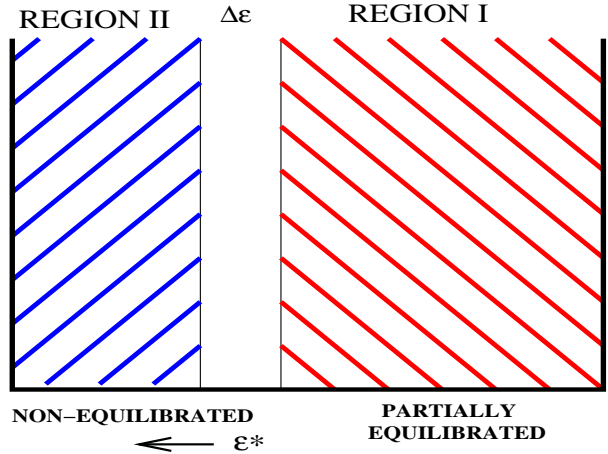}
\vskip 0.1in \caption{(Color online) Schematic picture showing the two
regions of modes. Region I ($\e>\e^*$) consists of partially
equilibrated modes at the temperature $\Theta(s)$. In region II
($\e<\e^*$) there are the non-equilibrated or diffusive modes. In
between the two regions there is a boundary region of width $\Delta
\e$ around the threshold $\e^*$ where modes start to equilibrate. As
time goes on, the threshold value $\e^*$ decreases with the width
$\Delta \e$ being proportional to $\e^*$. For sake of clarity the size
of the regions in the picture is not in scale (region II and $\Delta
\e$ should be tiny as compared to region I).\label{fig8}}
\end{center}
\end{figure}

To investigate how different energy modes relax to equilibrium we
introduce the mode-energy density \eq{des_energydens}, the
mode-magnetization \eq{des_modemag} and the mode-correlations and
mode-responses \eqq{mode_correl}{mode_response}. From the dynamical
equations equations \eq{des_modeenergy} we can prove the existence of
time-dependent threshold located at $\e^*(s)$ separating two
regions (I and II) characterized by the sign of the time derivative of
the mode-energy density
\eqqq{des_condition1}{des_condition2}{des_condition3}. Only the modes in
region I satisfy the equipartition relation at the effective temperature
$\Theta(s)$ \eq{des_equipartition}.  The threshold value $\e^*(s)$
decays with the age of the system with more modes entering region I as
the system approaches equilibrium. In Fig.\ref{fig8}
we show an schematic picture of the different mode regions in the model.

Furthermore, for each mode we can
define an equilibrium relaxation time \eq{des_eqtime} in terms of a
temperature dependent viscosity that diverges when $T\to 0$
\eq{des_gamma2}. Full correlation functions diverge if too many
diffusive modes are present, i.e. if $g(\e)$ vanishes too slowly in the
limit $\e\to 0$. For any equilibrium correlation function $C^{\rm
eq}(\tau)$ there exists a distribution $g(\e)$ that in the limit $\e\to
0$ reproduces the long-time decay of $C^{\rm eq}(\tau)$.

Mode-dependent effective temperatures can be also defined \eq{teffdes}
from mode-correlations and mode-responses. They agree with the previous
scenario. In region I modes are mutually equilibrated at the temperature
$\Theta(s)$ whereas in region II modes are out-of-equilibrium. The
latter show diffusive behavior and are characterized by an effective
temperature larger than $\Theta(s)$ and equal to $2\Theta(s)$ for $\e\to 0$. This last result is characteristic of purely diffusive
behavior ($X_{\infty}=1/2$, see \cite{des_felixsoll,des_felixcris,des_parisi}).
The effective temperature can also be computed in frequency
space~\eq{des_teffF} corroborating the results obtained in the
time-domain. It is noteworthy to mention that no frequency dependence is
found \eq{des_teffF2}. This result is in contradiction to what is
observed in experiments and might be consequence of the simplicity of
the model (see the discussion in the next section).

Finally, we have corroborated the scenario of mode-dependent effective
temperatures using the fluctuation relations recently proposed in the
framework of aging systems
\cite{des_felixnoeq2,des_crisfel,des_felixosc}. Interestingly, for both
partially equilibrated modes (region I) and diffusive modes (region II) the
fluctuation relation for the mode-magnetization gives the unique value
$\Theta(s)$. It is unclear to us whether this result is of
general validity or restricted to the DOM. It would be interesting to
put under scrutiny other types of models.

\section{CONCLUDING REMARKS}
\label{sec7}
In which ways can the DOM improve our current understanding of the
glassy state? A lacking piece in the glass transition puzzle is to
understand whether it is possible to identify effective parameters (such
as local temperatures or stresses) that reflect the nonequilibrium
relaxation toward equilibrium of macroscopic or bulk properties
(e.g. enthalpy, specific heat, volume, pressure, viscosity, thermal
conductivity,...). However, if dynamics and transport phenomena are
strongly heterogeneous then it is hard to understand how a description
of the macroscopic time evolution of bulk quantities can be achieved
from a few number of parameters as proposed by several phenomonological
approaches.

It would be very interesting to get experimental evidence that a unique
nonequilibrium or effective temperature can be identified in glass
formers in the aging state. Such temperature would constitute a
candidate parameter to describe many physical nonequilibrium properties
in the aging state. This point of view has been advocated a few years
ago by Nieuwenhuizen~\cite{theo}. The existence of a unique effective
temperature appears to be an inherent property of mean-field
systems. The DOM is yet another example of a mean-field system with a
well defined nonequilibrium or effective temperature (that we have
denoted by $\Theta(s)$). Nevertheless, the current experimental evidence
points toward the fact that dynamics is strongly heterogeneous and a
unique nonequilibrium temperature might not exist. Rather, the effective
temperature appears more as a local quantity that changes from region to
region inside the glass and which takes a value that depends directly on
the degree of mobility of the region.

\begin{figure}[hbp!]
\begin{center}
\includegraphics*[width=6cm,height=6cm]{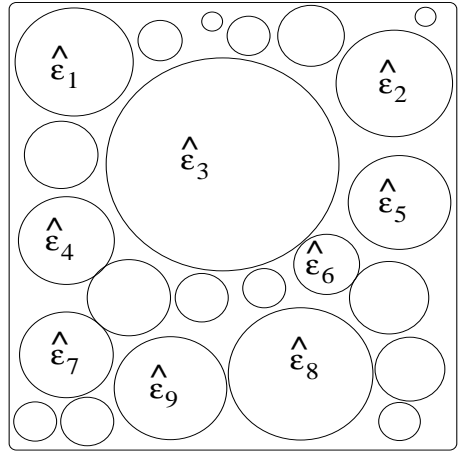}
\vskip 0.1in \caption{Schematic picture of a piece of glass made out of
different CRRs, each one characterized by a value of $\widehat{\e}$ (we
show 9 labelled regions in the picture). In such a mesoscopic
description regions are maybe of a few nanometers of spatial extension
and typical activation energies $K\Delta^2\widehat{\e}$ are of the order
of a few $k_BT$. $\widehat{\e}$ values fluctuate from region to
region. According to \eq{taurelax} dynamics is highly heterogenous with
relaxation times in local regions differing by many orders of
magnitude. The effective temperature \eq{des_infty} is also inversely
proportional to $\widehat{\eps}$ leading to an effective-temperature
field.
\label{mosaic}}
\end{center}
\end{figure}

The DOM is a mean-field model with cooperative glassy dynamics. Alone
it cannot explain the heterogenous behavior observed in glasses
because it does not incorporate the fact that different modes
intermittently switch for mobile to frozen behavior. The DOM describes
the dynamics of an isolated cooperatively rearranging region (CRR)
maybe of a few nanometers of spatial extension that includes a few
hundreds of atoms or molecules. The different energies $\e$ of the
vibrational modes in each CRR are determined by its volume (the number
of atoms it contains) and its surface through possible interactions
with nearby CRRs. The glass could then be viewed as a mosaic made out
of many CRRs each one described by a DOM with a given energy levels
distribution $g(\e)$. Such distribution strongly fluctuates among
CRRs. In the scenario where CRRs intermittently appear and dissapear
at a given rate, each time a new CRR is produced it carries a new
energy level distribution $g(\e)$. Consequently, at a given time each
region shows a different $\widehat{\e}$. According to \eq{taurelax},
fluctuations in $\widehat{\e}$ will lead to a great disparity of
relaxation times among CRRs and to the presence of slow and fast
regions. Also, the effective temperature will be inversely
proportional to $\widehat{\e}$ \eq{des_infty} leading to an
effective-temperature age-dependent field $\Theta(\widehat{\e},s)$ as
well as a $\Omega$-dependent effective temperature,
$T_{\textrm{eff}}^F(\Omega,\widehat{\e},s)$. The latter dependence of
the effective temperature on the probing frequency would agree with
the experimental results~\cite{des_cil2,des_cil1}. A schematic picture
of this scenario is shown in Fig.~\ref{mosaic}. This microscopic
scenario is similar to that proposed by Xia and Wolyness~\cite{XiaWol}
from the random first order transition theory for mean-field spin
glasses~\cite{des_bouch,KirkThir} applied to the glass transition. A
related scenario in terms of aggregating activated regions has been
also described by Crisanti and Ritort~\cite{CriRit02}. A more
elaborated and compelling description of the mosaic picture has been
recently proposed by Bouchaud and Biroli \cite{BouBir}.

Merging of random energy levels and cooperative dynamics in spatially localized
regions (CRRs) appears a promising route to provide a phenomenological
description of the glassy state. It remains to be seen whether such view
is capable to provide new predictions that can be tested in the
laboratory. This maybe difficult until we get a more clear picture and
better understanding about the nature of the heterogeneities uncovered by
the experimentalists. In turn, this might provide a definitive experimental evidence in
favour of the intuitive notion of CRRs put forward by Adam-Gibbs-Di Marzio many
years ago.

\appendix
\section{Dynamical equations for $Q_k$}

To solve the hierarchy of equations for $Q_k$ we start by computing
the evolution for the quantity $Q_2(t)$ \eq{des_q_2} which has appeared
in the dynamic evolution for the energy \eq{des_energy}. In order to do
that we compute the joint probability distribution of having a change
of $\delta E$ in the energy \eq{des_hamilt} and $\delta Q_2$ in
$Q_2$ \eq{des_q_2} for an elementary move \eq{des_element},

\beq P(\delta E,\delta Q_2) &=& \int_{-\infty}^\infty \delta
\left( \delta E - K\sum_i\e_i\left(\frac{r_ix_i}{\sqrt N}
+\frac{r_i^2}{2N}\right)\right)\nonumber\\
&\times&\delta \left( \delta Q_2 -
K\sum_i\e_i^2\left(\frac{r_ix_i}{\sqrt N}
+\frac{r_i^2}{2N}\right)\right) \left( \prod_i
\frac{dr_i}{\sqrt{2\pi\Delta_1^2}} \exp
\left(-\frac{r_i^2}{2\Delta_1^2}\right)\right). \eeq

This probability factorizes and can be written as follows:

\be P(x,y)=P(x)\frac{1}{\sqrt{4\pi J_2}}\exp\left(
\frac{-(y-J_1)^2}{4J_2}\right) ~~~,\ee \be J_1 = \frac{K\Delta^2
{\widehat{\e_2}}}{2}+( x-a_c) \frac{Q_3}{Q_2}~~,~~~~~ J_2 = K\Delta^2
\left( \frac{Q_3^2}{Q_2}+Q_4 \right), \ee where $x=\delta E$,
$y=\delta Q_2$ and $P(x)$ is given in \eq{des_pdeltae}. The other
quantities can be defined through the general expressions: \be
{\widehat{\e_k}}= \frac{1}{N}\sum_i \e^k~~,~~~~~ Q_k = \frac{K}{2N}\sum_i
\e^k x_i^2~~. \ee

From the joint probability we can derive the equation for
$Q_2(t)$ using \eq{des_metro}:

\begin{equation}
\frac{\partial Q_2}{\partial t} = \int_{-\infty}^0
dx\int_{-\infty}^\infty dy y P(x,y)+\int_0^\infty dx e^{-\beta
x}\int_{-\infty}^\infty dy y P(x,y)~~~.
\end{equation}

The integral over $y$ is just the quantity $J_1$ and does not depend
on $J_2$. Expressed in terms of the dynamical evolution of the energy
$\frac{\partial E}{\partial t}$ and the acceptance $A$ we find:

\be \frac{\partial Q_2}{\partial t} = \left( \frac{K\Delta^2
\widehat{\e_2}}{2}-\frac{a_c Q_3}{Q_2}\right) A(t) +
\left(\frac{Q_3}{Q_2} \right)\frac{\partial E}{\partial t}~~.  \ee

As we expect, the evolution of $Q_2$ depends also on the new
quantity $Q_3$. Therefore we have to generalize these computations
in order to obtain the whole set of coupled equations. The result
is:

\be \frac{\partial Q_k}{\partial t} = \left( \frac{K\Delta^2
\widehat{\e_k}}{2}-\frac{a_c Q_{k+1}}{Q_2}\right) A(t) +
\left(\frac{Q_{k+1}}{Q_2}\right) \frac{\partial E}{\partial t}~~.
\label{des_jer}~~~~~~ k\geq 2 \ee

These equations are, for a given probability distribution of the
disorder, the complete solution for the dynamical one-time
quantities.

At zero temperature only negative changes in energy are accepted.
The equations turn out to be a bit simpler:

\be\frac{\partial Q_k}{\partial t} = \frac{K\Delta^2
\widehat{\e_k}}{2} \textrm{erfc}(\alpha) -
\frac{Q_{k+1}}{Q_2}\sqrt{\frac{k\Delta^2Q_2}{\pi}}
\exp\left(-\frac{a_c^2}{4k\Delta^2Q_2}\right) \ee

We now outline the computations for the mode-energy density. As for
the quantities $Q_k$ we must first compute the joint probability of
having a change of $\delta E$ in the energy \eq{des_hamilt} and $\delta
w$ in $w(\e)$ \eq{des_energydens}, the result is:

\be P_\e(x,y)=P(x)\frac{1}{\sqrt{4\pi S_2(\e)}}\exp\left(
\frac{-(y-S_1(\e))^2}{4S_2(\e)}\right) ~~~,\label{eqfelix5}\ee where
now $x=\delta E$, $y=\delta w$ and $P(x)$ is given
in \eq{des_pdeltae}. The quantities $S_1(\e)$ and $S_2(\e)$ are given
by,

\be S_2(\e)= K\Delta^2 w(\e) - K\Delta^2\frac{(\e
w(\e))^2}{Q_2}~~~~,\ee \be S_1(\e)= \frac{\e
w(\e)}{Q_2}(x-a_c)+\frac{K\Delta^2 g(\e)}{2}~~~~.\ee

From the equation,
\begin{equation}
\frac{\partial w(\e,t)}{\partial t} = \int_{-\infty}^0
dx\int_{-\infty}^\infty dy y P(x,y)+\int_0^\infty dx e^{-\beta
x}\int_{-\infty}^\infty dy y P(x,y)~~~,
\end{equation} we get: \be \frac{\partial w(\e,t)}{\partial t}= \left(
\frac{\e w(\e,t)}{Q_2(t)}\right) \frac{\partial E}{\partial
t}+\left( \frac{K\Delta^2 g (\e)}{2}-\frac{a_c\e
w(\e,t)}{Q_2(t)}\right) A(t)~~. \label{des_modeenergy2} \ee

\section{Equilibrium behavior of the correlations and responses}

In this appendix we will solve the dynamical equations for the
mode-correlation and the mode-response functions and analyze their
equilibrium behavior. First of all, let us define:

\be f(t)=-\left( \frac{\partial E}{\partial t}-a_cA(t)
\right)~~.\label{des_function} \ee

In terms of this function the solution for the
mode-correlation \eq{des_corrvst} is given by:

\be C(\e,t,s)= C(\e,s,s) \exp\left(
-\e\int_s^t\frac{f(x)}{2Q_2(x)}dx\right)~~~.\label{des_corrgen}\ee

Where the initial condition for the mode-correlation is:
$C(\e,s,s)= \frac{2}{K} w(\e,s)$. In a more compact way we can
write:

\be C(\e,t,s)=\frac{2}{K} w(\e,s)\exp\left(-\int_s^t F(\e,x)dx
\right)~~~~~~~{\rm{with}}~~~~~ F(\e,x)= \e
\frac{f(x)}{2Q_2(x)}~~.\label{des_corrf}\ee

In general we can define a time-dependent relaxation time,

\be \tau(\e,t,s)=
\frac{(t-s)}{\e}\left(\frac{1}{\int_s^t\frac{f(x)}{2Q_2(x)}dx}
\right)~~~~,\label{des_tdtime2}\ee and the correlation function
can be written as:

\be C(\e,t,s)=\frac{2}{K}w(\e,s)\exp\left[ -(t-s)/\tau
(\e,t,s)\right]~~~~.\ee

In equilibrium $Q_2^{\rm{eq}}$ is given by \eq{des_q2eq} and
$f^{\rm{eq}}$ \eq{des_function} is given by: \be
f^{\rm{eq}}=a_c \textrm{erfc}(\alpha^{\rm{eq}})~~~~,\ee where $
\alpha^{\rm{eq}}$ is given in \eq{eqIIIerfc}. Therefore, the
mode-correlation in equilibrium has an exponential behavior:

\be C^{\rm{eq}}(\e,t,s)= \frac{2w^{\rm{eq}}}{K}\exp\left[
  -(t-s)/\tau^{\rm{eq}} (\e)\right]~~~~,\label{des_correqap}\ee where
$w^{\rm{eq}}$ is given in \eq{des_eqcond} and the characteristic
equilibration time of the energy mode $\e$ is: 
\be \tau^{\rm{eq}} (\e)
= \frac{1}{K\e}\left(\frac{2T}{\Delta^2
  \textrm{erfc}(\alpha^{\rm{eq}})}\right)~~~~~.\label{des_eqtime}\ee
As we have mentioned before the characteristic relaxation time of all
the oscillators belonging with energy mode $\e$ is inversely
proportional to the value of $\e$. Therefore, the higher is the value
of $\e$ the shorter is the characteristic time-scale of the
oscillator. Therefore we can define the friction coefficient $\gamma$
using the relation $\tau^{\rm{eq}}(\e)=\frac{\gamma}{K\e}$ for an overdamped
harmonic oscillator in thermal equilibrium:

\be \gamma =\left(\frac{2T}{\Delta^2
  \textrm{erfc}(\alpha^{\rm{eq}})}\right)~~~~,\label{des_gamma}\ee
  which, for low enough temperatures behaves, \be \gamma
  =\sqrt{\frac{TK\widehat{\e}\pi}{2\Delta^2}}\exp \left[
  \frac{K\Delta^2 \widehat{\e}}{8}~~\frac{1}{T} \right]~~~~.\label{des_gamma2}\ee

Therefore, the divergence of the relaxation time is of an
activated type just as in the model without disorder.

We can also derive the equilibrium relaxation time from the
response function:
\be G(\e,t,s)=\frac{F(\e,s)g(\e)}{\e K}
\exp\left(-\int_s^t F(\e,x)dx \right)\theta(t-s)~~,\label{des_respgen}
\ee
with $F(\e,x)$ given in \eq{des_corrf}. In equilibrium the mode-response becomes:

\be G^{\rm{eq}}(\e,t,s)=\frac{g(\e)}{\e\tau^{\rm{eq}}(\e)K} \exp\left[
  -(t-s)/\tau^{\rm{eq}} (\e)\right]\theta(t-s)~~~~.\ee

Now we can check that in equilibrium FDT is verified :

\be \left(\frac{\frac{\partial C(\e,t,s)}{\partial s}}{G(\e,t,s)}
\right)= \frac{2\e w(\e)}{g(\e)} = T~~.\ee

\section{Mode-magnetization rates}

In this section we derive the main results for the mode-magnetization
rates. The joint probability of having a change in the mode
magnetization \eq{des_modemag} $\delta M_\e$ and in the energy
$\delta E$ can be
computed as in the original model \cite{des_art1,des_felixosc}. The result is:

\be P^h(\delta E,\delta M_\e)=\frac{1}{4\pi\sqrt{S_2(\e)S_4(\e)}}\exp\left(
\frac{-(\delta E-a_c)^2}{4S_2(\e)}\right)\exp\left(\frac{-(\delta M_\e
-\frac{S_3(\e)}{2S_2(\e)}(\delta E - a_c))^2}{4S_4(\e)}\right)~~~~,\label{eqc1}\ee where
we have defined the quantities:

\be S_2(\e) = K\Delta^2Q_2-h\e K\Delta^2 M + {\cal{O}}(h^2)~~~~, \ee
\be S_3(\e)= \e M K\Delta^2 - h\Delta^2 g(\e)~~~~,\ee \be
S_4(\e)=\frac{\Delta^2 g(\e)}{2}+ {\cal{O}}(h^2)~~~~. \ee

At $T=0$ the probability that an attempted change $\delta M_\e$ is
accepted is given by:

\be W^h(\delta M_\e) = \int_{-\infty}^0 P^h(\delta E,\delta M_\e)
d(\delta E)~~~~.\ee

Assuming that the value of the magnetization is zero (condition of
neutrality \cite{des_felixosc,des_sollteff1}) we find:

\be W^h(\delta M_\e) = W^0(\delta M_\e)\exp\left( \delta M_\e
h\frac{\widehat{\e}}{4Q_2} \right)~~~~,\label{des_probm}\ee where
\be W^0(\delta M_\e) = \frac{1}{4\pi\sqrt{S_2(\e)S_4(\e)}}\exp\left( -
\frac{(\delta M_\e)^2}{4S_4(\e)}\right)\int_{-\infty}^0 \exp\left(
\frac{-(\delta E-a_c)^2}{4S_2(\e)}\right) d(\delta E)~~~~,\ee and
therefore,

\be \frac{W^h(\delta M_\e)}{W^h(-\delta M_\e)}=\exp \left( \delta M_\e h
\frac{\widehat{\e}}{2Q_2}\right)~~~~.\ee

\section{The effective temperature in Fourier space}

The effective temperature in Fourier space is defined
as \eq{eqfelix3}:
\be T_{\textrm{eff}}^F(\eps,\Omega,s)=\Omega
\frac{S(\eps,\Omega,s)}{\widehat{\chi}''(\eps,\Omega,s)} ~~~~.\ee
In our case, the real part of the power spectrum is given by,
\be S(\e,\Omega,s)=\frac{2}{K}w(\e,s) \int_s^\infty dt \cos (\Omega
t) \exp[-(t-s)/\tau(\e,t,s)]~~~~,\label{eqfelix3b}\ee where we have
used \eq{des_corrtau}. The susceptibility is,
\be \chi(\e,t,s)=\int_s^t G(\e,t,\tau) d\tau=\frac{g(\e)}{\e
K}\left(1-\exp[-(t-s)/\tau(\e,t,s)]\right)~~~~,\ee where we have
used \eq{des_resptau}. The complex part of the response is defined
as \eq{eqfelix4}:
\be \widehat{\chi}(\eps,\Omega,s)=i\Omega \int_s^\infty dt \chi
(\eps,t,s)e^{-i\Omega
t}=\widehat{\chi}'(\eps,\Omega,s)-i\widehat{\chi}''(\eps,\Omega,s)~~~~ .
\ee
Taking the imaginary part we get,
\be \widehat{\chi}''(\e,\Omega,s)=\frac{g(\e)}{\e K} \int_s^\infty dt
\cos \Omega t \exp[-(t-s)/\tau(\e,t,s)]~~~~.\ee
The final result is:
\be T_{\textrm{eff}}^F(\eps,\Omega,s)=
\frac{2\e}{g(\e)}w(\e,s)~~~~.\label{des_fourierteff}\ee

Note that the effective temperature in Fourier space does not depend
on the frequency $\Omega$.  This result is consequence of the
identical functional decay of the correlation and response functions
\eqq{des_corrf}{des_respgen} which leads to nearly identical
expressions for the power spectrum \eq{eqfelix3b} and the
frequency-dependent susceptibility \eq{eqfelix4}. In this respect this
result is not generic of mean-field models which instead do show a
dependence of the effective temperature
$T_{\textrm{eff}}^F(\eps,\Omega,s)$ on the probing frequency $\Omega$.


\begin{center}
\textbf{ACKNOWLEDGMENTS}
\end{center}

We wish to thank P. Sollich for interesting comments and suggestions. A. G. has 
been supported by the University of Pompeu Fabra. F.  R is
supported by the Ministerio de Ciencia y Tecnolog\'{\i}a in Spain
{\bf{BFM2001-3525}}, SPHINX (ESF) program, STIPCO
\textbf{HPRN-CT-2002-00319}, and the Distinci\'o de la Generalitat de
Catalunya.

\end{document}